\documentclass[aps,a4paper,showpacs,preprintnumbers,amsmath,amssymb,
  floatfix]{revtex4}
\usepackage{dcolumn}
\usepackage{bm}
\usepackage{amsmath,amssymb}
\usepackage{booktabs,subfigure}
\usepackage{graphicx,psfrag}
\usepackage{epsfig}
\usepackage{color} 
\usepackage{rotating}
\usepackage{slashed}
\usepackage{geometry}
\usepackage{rotating}
\usepackage{subfigure}
\usepackage{caption}

\usepackage{pstricks}
\usepackage{color}
\allowdisplaybreaks[3]
\newcommand{\be}{\begin{equation}}
\newcommand{\ee}{\end{equation}}
\newcommand{\bea}{\begin{eqnarray}}
\newcommand{\eea}{\end{eqnarray}}
\newcommand{\cw}[1][{}]{\ensuremath{\cos^{#1} \theta_{W}}}
\newcommand{\sw}[1][{}]{\ensuremath{\sin^{#1} \theta_{W}}}

\numberwithin{equation}{section} 
\usepackage{epsfig}
\usepackage{dcolumn} 
\usepackage{bm} 

\def\gsim{\lower0.5ex\hbox{$\:\buildrel >\over\sim\:$}}
\def\lsim{\lower0.5ex\hbox{$\:\buildrel <\over\sim\:$}}

\begin{document}

\title{Invisible decays of the heavier  Higgs boson in the minimal supersymmetric standard model}

\vskip 2cm

\author{B.Ananthanarayan, ~Jayita Lahiri,  ~P. N. Pandita}

\affiliation{ Centre for High Energy Physics, 
                  Indian Institute of Science, Bangalore 560 012, India}


\thispagestyle{myheadings}


\vskip 5cm

\begin{abstract}
\noindent
We consider the possibility that the heavier CP-even Higgs boson~($H^0$) 
in the minimal supersymmetric standard model (MSSM)
decays invisibly into neutralinos in the light of the recent discovery of 
the 126 GeV resonance at the CERN Large Hadron Collider (LHC). 
For this purpose  we  consider the minimal supersymmetric standard model
with universal, non-universal and arbitrary boundary conditions on the 
supersymmetry breaking gaugino mass parameters at the grand unified scale. Typically,
scenarios with universal and nonuniversal gaugino masses 
do not allow invisible decays of the lightest
Higgs boson~($h^0$), which is identified with the $126$ GeV 
resonance, into the lightest neutralinos in the MSSM.  
With arbitrary gaugino masses at the grand unified scale such an invisible decay is possible.
The second lightest Higgs boson can decay into various 
invisible final states for a considerable region of the MSSM parameter
space with arbitrary gaugino masses as well as with the gaugino masses 
restricted by universal and nonuniversal boundary conditions at the 
grand unified scale.The possibility 
of the second lightest Higgs boson of the MSSM decaying into invisible channels 
is more likely for arbitrary gaugino masses at the grand unified scale.
 The heavier Higgs boson decay into lighter particles leads to the intriguing possibility that the 
entire Higgs boson spectrum of the MSSM
may be visible at the LHC even if it decays invisibly, during the searches for
an extended Higgs boson sector at the LHC. 
In such a scenario the nonobservation of the extended Higgs sector of the
MSSM may carefully be used to rule out regions of the MSSM parameter space at the LHC.

\end{abstract}

\pacs{12.60.Jv, 14.80.Da, 14.80.Ly, 14.80.Nb}
\maketitle

\section{Introduction}
\label{sec:intro}
With the discovery~\cite{arXiv:1207.7214,ATLAS:2013mma, 
arXiv:1207.7235,Chatrchyan:2013lba}  
of  a neutral state around a mass of $126$ GeV at the 
CERN Large Hadron Collider~(LHC), a new era 
has begun in our quest for the understanding of the fundamental constituents
of matter and the forces between them. Although,  the properties of the 
discovered state are  consistent with the properties of the standard model~(SM) 
Higgs boson, it also opens up a window for searches of new physics. It is well
known that the SM Higgs sector suffers from the naturalness and hierarchy
problems, thereby rendering a light Higgs boson technically unnatural.
The most popular, and well motivated,
extension of the SM which renders a light Higgs boson technically natural include 
supersymmetric models~\cite{Wess:1992cp}, of which the minimal supersymmetric 
standard model~(MSSM) is perhaps the most economical, 
and hence compelling~\cite{Nilles:1983ge}.  
In the MSSM the Higgs spectrum is richer as compared to the  Higgs sector 
of the SM, and consists
of two Higgs superfields~($H_1$ and $H_2$). After spontaneous symmetry 
breaking the model contains two CP even Higgs bosons~($h^0$, $H^0$; $M_{h^0} < M_{H^0}$),
one CP odd Higgs boson~($A^0$), and two charged states~($H^\pm$). 
At the tree level the  Higgs sector of MSSM is rather constrained and is
described by two parameters, usually taken to be the mass of $A^0$~($M_A$) 
and the ratio of the vacumm expectation values of the two Higgs fields,
$\tan\beta \equiv  \langle H_2^0 \rangle/ \langle H_1^0 \rangle$. 
Discovery of more than one Higgs boson
at the LHC would point towards an extension of the SM, of which 
supersymmetry as embodied in the minimal supersymmetric
standard model is a leading candidate.
It would, therefore, be important to discover an extended Higgs sector
at the LHC, if it exists. However, from the point of view of supersymmetry,
it is crucial to discover the supersymmetric partners of the SM states
as  predicted by the MSSM such as squarks, gluinos and sleptons,
as well as neutralinos and charginos.  
In the absence of any signal for supersymmetric particles,
it would, therefore,  be appropriate to  ask
the question whether non  observation of an extended Higgs sector
would imply that the new physics is at a much higher scale.  
This question is intimately 
connected with the decay patterns of the Higgs bosons of the MSSM.
It is possible that the Higgs bosons of the MSSM may decay into
some of the lighter particles of the supersymmetric spectrum at a rapid rate. 
This, in particular, would include light neutralino pairs. This decay can be an important decay channel in certain regions of the MSSM parameter space. Hence we need to study the invisible decay of the second lightest Higgs boson extensively.  Another aspect of this issue is that even though MSSM could
be ruled out, there are appealing alternatives, namely the nonminimal 
or next-to-minimal supersymmetric standard model~(NMSSM)
whose Higgs~\cite{Fayet:1974pd, Nilles:1982dy, Ellis:1988er, Drees:1988fc, Pandita:1993hx,
Pandita:1993tg, Ellwanger:1993hn, Elliott:1993uc}
and neutralino~\cite{Pandita:1994ms, Pandita:1994vw, Choi:2001ww} sectors are richer than 
the MSSM, thereby increasing the possibility of
an invisibly decaying Higgs boson. Furthermore, in the NMSSM there is a 
possibility of a low mass pseudoscalar in the Higgs spectrum. Thus,
there is a distinct possibility of the scalar Higgs boson(s) decaying 
invisibly into these pseudoscalars~\cite{Ananthanarayan:2013fga}.

The invisible Higgs decay width has been constrained by
various groups by performing fits of the signal strengths in various search 
channels using the latest LHC Higgs
data~\cite{Belanger:2013,Djouadi:2013mn}. Direct searches for invisible decaying Higgs 
produced in association with a Z boson has been carried out by the
ATLAS~\cite{ATLAS:2014av,ATLAS:2013av} and CMS~\cite{CMS:2013pr} collaborations at the 
LHC and they have excluded branching ratio of more than 65\% and 75\%, respectively, 
with 95\% CL. The CMS collaboration~\cite{CMS:2013sj} has also carried out a similar 
search for invisible branching ratio of the Higgs boson produced in the vector boson 
fusion process and put an upper limit of 69\% on the invisible branching ratio of the 
lightest Higgs boson. All these searches have in turn put constraints on the MSSM 
parameter space. Although specific regions of the MSSM parameter space have been ruled out 
by these experiments, there is still a large portion of the MSSM parameter space which 
remains unexplored. Recently, the question of the invisible decays of the lightest Higgs 
boson in the context of the MSSM and the NMSSM have been discussed in
detail~\cite{Ananthanarayan:2013fga, Pandita:2014nsa}. In this paper we 
carry this investigation further by analyzing the decay patterns and the 
invisible decays of the heavier Higgs boson~($H^0$) of the MSSM. 
For this purpose we shall identify the state observed near $126$~GeV 
at the CERN LHC with the lightest Higgs boson~($h^0$)
of the MSSM.  We shall systematically study the scenarios under which such
a possibility can  arise and discuss different  aspects of these scenarios.

The plan of the paper is as follows. In Section II we discuss the relevant features 
of the Higgs sector of the MSSM, and enumerate the two regions, the decoupling and the 
non-decoupling regions, of its parameter space.
In Section III we discuss the neutralino sector and the neutralino mass matrix of the MSSM
with two different kinds of grand 
unified theory (GUT) boundary conditions on its  parameters, 
namely universal and non-universal boundary conditions and the composition of neutralinos. 
In Section IV we summarize the analytical expressions for 
the decay of second lightest Higgs boson 
to neutralinos in the MSSM with an appropriate choice of the parameter space, 
taking all the experimental contraints into account. In Section V we present 
our calculations and numerical results for the invisible decay of the second 
lightest Higgs boson and comment on how some of them may be rendered visible(quasi-invisibility). Finally 
in Section VI we summarize our results and conclusions.

\bigskip

\section{The Higgs sector of minimal supersymmetric standard model}

\bigskip

The ATLAS~\cite{arXiv:1207.7214,ATLAS:2013mma} 
and CMS~\cite{arXiv:1207.7235,Chatrchyan:2013lba}  
experiments at the CERN LHC have  independently
observed a resonance, whose properties are consistent with
the SM Higgs boson. The ATLAS experiment after collecting data at an
integrated luminosity of $4.8$ fb$^{-1}$ at $\sqrt{s}$ = $7$ TeV 
in $2011$ and $5.8$ fb$^{-1}$ at $\sqrt{s}$ = $8$ TeV in $2012$ 
confirmed the evidence for the production of a neutral boson with a measured 
mass of $126.0 \pm 0.4(\rm stat) \pm 0.4(\rm sys)$ GeV,  with a significance of 
$5.9$ $\sigma$. The CMS experiment after collecting 
$5.1$ fb$^{-1}$ at $7$ TeV 
and $5.3$ fb$^{-1}$ at $8$ TeV reported an evidence of a neutral boson 
at $125.3 \pm 0.4 (\rm stat.) \pm 0.5 (\rm syst.)$ GeV with a   
significance of $5.8$ $\sigma$. As mentioned in the Introduction,
within the framework of the MSSM, 
we shall identify this  resonance with the lightest Higgs boson of the model. 
We recall that the
tree-level Higgs boson masses are determined by CP-odd Higgs boson mass $M_A$ 
and  $\tan \beta$. 
Requiring the production cross section of 126 GeV Higgs boson decaying to two 
photons agrees with the one observed at the CERN LHC, 
divides the MSSM Higgs parameter space into two 
distinct regions~\cite{Li:2014qv}:

\begin{itemize}

\item{ The non-decouplng regime where $M_A \lsim 130$ GeV. In this region the heavier 
CP-even state $H^0$ 
is SM-like, and the light CP-even Higgs state $h^0$ and the CP-odd Higgs state $A^0$ 
are almost degenerate in mass~\cite{Pandita:1984hf} and close to $M_Z,$ 
while the charged Higgs bosons are nearly degenerate with $H^{0}$~\cite{Heinemeyer:2012}. 
The LHC phenomenology for this sector has been discussed in the 
past~\cite{Boos:2002,Boos:2004,Han:2012}.}

\item{The decoupling limit where  $M_A \gsim 300$ GeV.  In this region 
the light CP-even Higgs boson $h^0$ is SM-like, and all the other physical 
Higgs boson(s) are heavy and almost degenerate with $A^0$~\cite{Haber:2014}. 
The decoupling properties of the
MSSM Higgs sector are not special to supersymmetry.
Rather, they are a generic feature of non-minimal Higgs sectors.}

\end{itemize}

The non-decoupling scenario, which  leads to a light SM-like Higgs, 
is highly constrained. Thus, the decoupling 
regime is a more viable  scenario as far as the MSSM Higgs 
search results are concerned. Therefore, we shall consider only this scenario
in this paper. For the non-decoupling regime of supersymmetric models, 
see, e.g.~\cite{Pandita:2014nsa}.

Having summarized the experimental results and the different scenarios, we now
summarize the aspects of the MSSM Higgs sector that are relevant to our
discussion. The masses $M_{h/H}$  and the mixing angle $\alpha$
of the neutral CP-even Higgs states are well known. These can be 
written as~\cite{Djouadi:2013qv}
\begin{eqnarray}
M_{h/H}^2& = &\frac{1}{2}\ ( M_{A}^2+M_{Z}^2+ \Delta {M}_{11}^2+ 
\Delta {M}_{22}^2  \mp  \sqrt{ M_{A}^4+M_{Z}^4-2 M_{A}^2 M_{Z}^2 
c_{4\beta} +C}\ ), \\
\tan \alpha & = &\frac{2\Delta {M}_{12}^2 - (M_{A}^2 + M_{Z}^2) s_{\beta}}
{ \Delta {M}_{11}^2 -  \Delta {M}_{22}^2 + (M_{Z}^2-M_{A}^2)
c_{2\beta} + \sqrt{M_{A}^4 + M_{Z}^4 - 2 M_{A}^2 M_{Z}^2 c_{4\beta} + C}},
\label{eqn1}
\end{eqnarray}
where
\begin{eqnarray}
C & = & 4 \Delta {M}_{12}^4 + ( \Delta {M}_{11}^2 - 
\Delta {M}_{22}^2)^2 -  
 2 (M_{A}^2  -  M_{Z}^2)( \Delta {M}_{11}^2  -  \Delta M_{22}^2) 
 c_{2\beta} 
- 4 (M_{A}^2  + M_{Z}^2)  \Delta {M}_{12}^2 s_{2\beta},
\label{eqn2}
\end{eqnarray}
and
$c_{4\beta} \equiv \cos 4\beta, c_{2\beta} \equiv \cos 2\beta,
c_{\beta} \equiv \cos \beta,$ and $s_{\beta} \equiv \sin\beta.$
Furthermore,  $\Delta {M}_{ij}, i, j = 1,2$ quantify the radiative
corrections to the CP-even Higgs boson mass matrix.

The dominant radiative corrections arising
from the top-stop sector are contained in $\Delta {M}_{22}.$ Identifying
the resonance discovered at $126$ GeV  with $h^0,$
one can, then,  write the dominant radiative correction to the Higgs boson mass 
matrix in terms of the mass $M_h$ of $h^0$ as 
\begin{equation}
\Delta {M}^{2}_{22}= \frac{M_{h}^2(M_{A}^2  + M_{Z}^2 -M_{h}^2) - M_{A}^2 
M_{Z}^2 c^{2}_{2\beta} } { M_{Z}^2 c^{2}_{\beta}  +M_{A}^2 s^{2}_{\beta} 
- M_{h}^2},
\label{eqn3}
\end{equation}
where we have used the approximation
$\Delta {M}_{22} \gg \Delta {M}_{11}, \Delta {M}_{12},$ and where $M_A$ is the 
mass of the pseudoscalar Higgs boson $A^0.$ 
We note that there is another solution for $\Delta M_{22}^2,$ which is
unphysical~\cite{Maiani:2012qf}, and is, therefore, not relevant. 

\begin{figure}
\begin{minipage}[b]{0.45\linewidth}
\centering
\includegraphics[width=6cm, height=5cm]{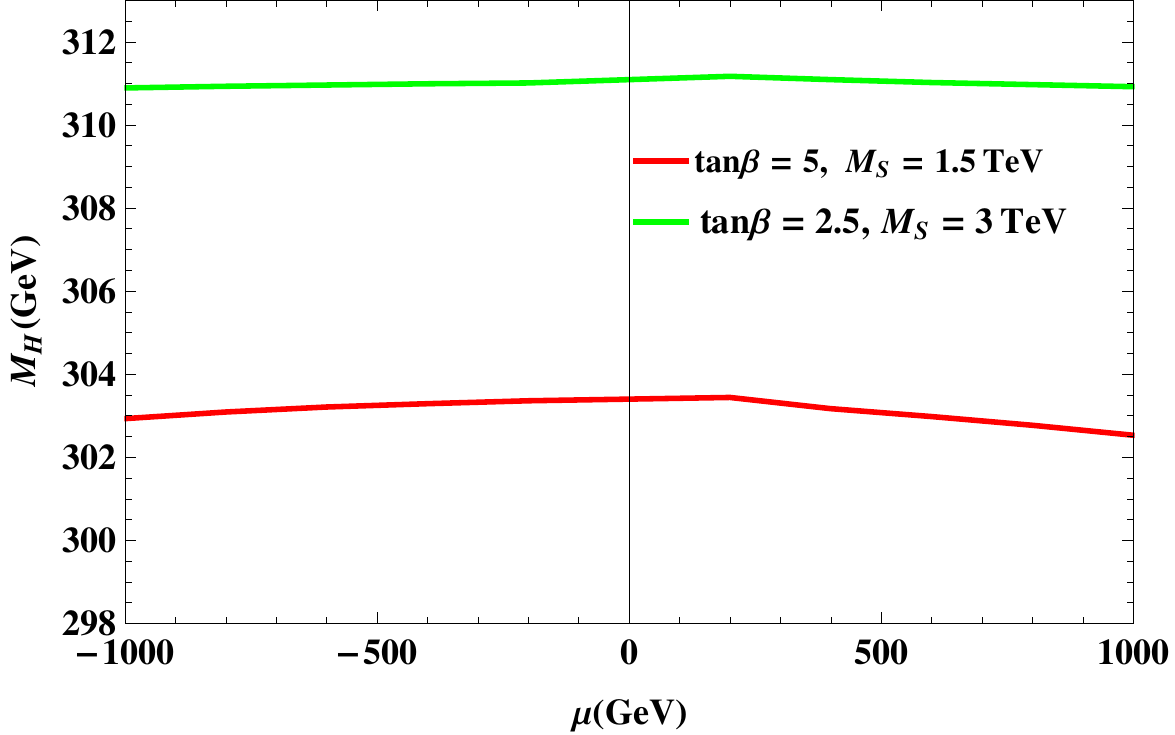}
\caption{$M_H$ as a function of $\mu$ for $M_A$ = 300 GeV
for two different values of $\tan\beta$ and the supersymmetry breaking
scale $M_S.$}
\label{fig:mu_MH_300}
\end{minipage}
\hspace{0.4cm}
\begin{minipage}[b]{0.45\linewidth}
\centering
\includegraphics[width=6cm, height=5cm]{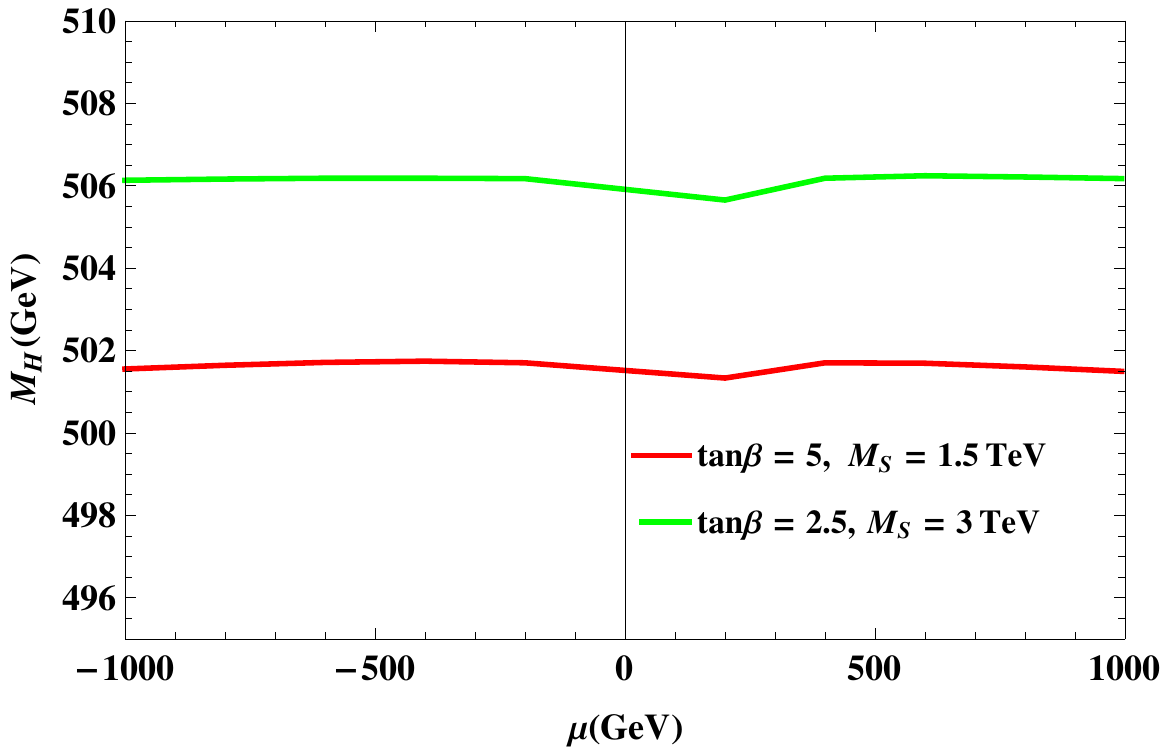}
\caption{$M_H$ as a function of $\mu$ for $M_A$ = 500 GeV
for two different values of $\tan\beta$ and the supersymmetry breaking
scale $M_S.$}
\label{fig:mu_MH_500}
\end{minipage}
\end{figure}

With this approximation
we can write the mass of $H^0$ and the mixing angle $\alpha $ 
in terms of $M_A$ and $\tan\beta$ as
\begin{eqnarray}
M_H^2 & = & \frac{(M_A^2 + M_Z^2 -M_h^2)( M_Z^2c_{\beta}^2 + M_A^2 s_{\beta}^2)
              - M_A^2 M_Z^2 c_{2\beta}^2}
	      {M_Z^2 c_{\beta}^2 + M_A^2 s_{\beta}^2 -M_h^2}, \\
\alpha & = & - \tan^{-1} \left(\frac{(M_Z^2 + M_A^2) c_{\beta} s_{\beta}}
                              {M_Z^2 c_{\beta}^2 + M_A^2 s_{\beta}^2 - M_h^2}
			      \right).
\label{eqn4}
\end{eqnarray}
\noindent
Thus, in principle one can calculate the mass of $H^0.$ However, in actual 
practice the mass of $H^0$ will depend on other parameters of the model,
which include the supersymmetry conserving Higgs bilinear parameter $\mu$, 
and the supersymmetry breaking scale $M_S$.  In order to obtain a handle on the 
behavior of solutions of interest to us,
we perform a  numerical scan using the package CalcHEP~\cite{Belyaev:2012qa}. 
For our study we use a set of  inputs which are consistent with
the known experimental constraints, and also which have the possibility
of leading to spectra that may be visible in the near future.  Thus, we
are guided by the principle of search for SUSY in upcoming experiments
and choices of parameters that continue to make low energy SUSY a
viable option to address the naturalness and hierarchy problems of the SM.
We have calculated the dependence of $M_H$ on $\mu$ for different 
values of $\tan \beta$ and the supersymmetry breaking scale $M_S$. 
In Figs.~\ref{fig:mu_MH_300} and \ref{fig:mu_MH_500} 
we show the dependence of $M_H$ on $\mu$ 
for values of $M_A$ = 300 GeV and $M_A = 500$ GeV.
From these two figures we conclude that $M_H$ does not vary significantly 
as a function of $\mu.$ 
For fixed values of $\tan \beta$ it has a weak dependence on 
$M_S$. However, $M_H$ has a significant dependence 
on $M_A$ and $\tan \beta$, and can be described fully in terms of these two parameters 
when we use the fact that $M_h$ is in the range 123-129 GeV. This feature had been 
pointed out earlier in Refs.~\cite{Maiani:2012,Maiani:2012rc,Djouadi:2013az}. Our results
agree with these authors. In the present work these results have
been obtained numerically by using as input the parameters 
$M_S$, $\tan \beta$ and $\mu$ as detailed above, 
and adjusting the scalar trilinear coupling $A_t$ 
so as to have  $M_h$ in the range 123-129 GeV. 
We have checked that the results of our calculations
are in good agreement with those of ~\cite{Djouadi:2013qv}.

\section{The neutralino sector of the MSSM with GUT boundary conditions}
\label{sec:gut_boundary}

 In order to study the invisible decays of $H^0$ to neutralinos, we consider the 
neutralino sector of the MSSM in some detail. For this purpose we consider the 
neutralino mass matrix of the model and the implications of the GUT boundary conditions
on the neutralino spectrum. 
This analysis will lead to a general understanding of the nature of
mixing between the gaugino and Higgsino states, as well as those of
the physical neutralino states after electroweak symmetry breaking.

We recall that the neutralinos are an admixture of the 
fermionic partners of the two Higgs doublets, $H_1$ and $H_2$,
and the fermionic partners of the neutral gauge bosons. When the 
electroweak symmetry is broken, the physical mass eigenstates 
are obtained from the diagonalization of the neutralino 
mass  matrix~\cite{Bartl:1989ms,Haber:1984rc,Guchait:1993rc}
\begin{eqnarray}
\label{mssmneut}
M_{\mathrm{MSSM}} =
\begin{pmatrix}
M_1 & 0   & - m_Z \sw \cos\beta & \phantom{-}m_Z\sw \sin\beta \\
0   & M_2 & \phantom{-} m_Z \cw \cos\beta  & -m_Z \cw\sin\beta \\
 - m_Z \sw \cos\beta &\phantom{-} m_Z \cw \cos\beta  & 0 & -\mu\\
\phantom{-}m_Z\sw \sin\beta& -m_Z \cw\sin\beta & -\mu & 0
\end{pmatrix},
\label{matrix1}
\end{eqnarray}
\noindent where $M_1$ and $M_2$ are the $U(1)_Y$ and 
the $SU(2)_L$ soft supersymmetry breaking gaugino
mass parameters, $\mu$ is the Higgs(ino) mass parameter, $M_Z$ is the
$Z$ boson mass, $\theta_W$ is the weak mixing angle and 
$\tan\beta \equiv v_2/v_1$ is the ratio of the vacuum expectation values 
of the neutral components of the 
two Higgs doublet fields $H_1$ and $H_2$. The gaugino mass parameters $M_1$ and $M_2$  may 
have some relation between them, or they can be completely arbitrary. 
If we assume that the MSSM is embedded in a  grand unified theory, then the 
boundary conditions at the GUT scale imply  a definite relation  between the gaugino masses,
which would imply a relation between them at the weak 
scale following the renormalization group evolution. Here we will consider  
two types of boundary conditions on $M_i$ that follow from embedding of 
MSSM in a grand unified theory, namely the  universal boundary conditions 
and the nonuniversal boundary conditions.

\subsection{Universal boundary condition}
\label{subsec:universal-gaugino-masses}

In the MSSM, with universal gaugino masses at the grand unified scale, 
usually referred to as mSUGRA, the soft supersymmetry breaking
gaugino mass parameters $M_i$ and the corresponding gauge couplings
$\alpha_i (i = 1, 2, 3)$  satisfy the  
boundary condition 
\bea
M_1 & = &  M_2 = M_3 = m_{1/2}, \\
\label{gauginogut}
\alpha_1 & = & \alpha_2 =  \alpha_3 = \alpha_G,  \label{gaugegut}
\eea
at the grand unified scale $M_G.$
Using the one-loop 
renormalization group equations~\cite{Martin:1993ft}
for the gaugino masses 
and the gauge couplings  leads to the ratio
\begin{equation}
M_1 : M_2 : M_3 \simeq 1 : 2 : 7.1,
\label{msugra0}
\end{equation}
\noindent
for the soft gaugino masses at the electroweak scale $M_Z$. 
In the following, for definiteness,  we shall consider the
value of $\tan \beta$ = 10.
If one is interested in Higgs boson decaying invisibly
into light neutralinos, it is  appropriate to consider
a light or massless eigenstate of the mass matrix (\ref{matrix1}).
It has been shown earlier~\cite{Ananthanarayan:2013fga,Pandita:2014nsa}
that with the gaugino mass parameters satisfying the universal boundary
conditions at the GUT scale, it is not possible to satisfy the 
masslessness condition for the lightest neutralino
following from the determinant of the  mass matrix (\ref{matrix1})
and the experimental constraints as implied by
the LEP experiments~\cite{ALEPH:2005ab}.
Thus, a massless neutralino is ruled out with universal boundary conditions
(\ref{gauginogut}) at the GUT scale.


\subsection{Non-universal boundary condition}

Universal soft supersymmetry breaking gaugino
masses are not the only possibility in a grand unified theory. In fact,
non universal boundary conditions for the soft gaugino masses can 
arise quite naturally  in $SU(5), SO(10)$ and $E_6$ 
supersymmetric grand unified theories.

In grand unified supersymmetric models, gaugino masses 
are generated by a non-singlet chiral superfield $\Phi^n$ that appears 
linearly in the gauge kinetic function 
$f(\Phi)$, which is an analytic function of the 
chiral superfields $\Phi$ in the theory\cite{Cremmer:1982wb}.
The gaugino masses are generated from the coupling of 
the field strength superfield $W^a$ with $f(\Phi)$,
when the auxiliary part $F_\Phi$ of a chiral superfield $\Phi$  in 
$f(\Phi)$ gets a VEV.
When $F_{\Phi}$ gets a VEV $<F_{\Phi}>,$ we obtain the 
Lagrangian containing the gaugino masses 
\begin{equation}
{\cal L}_{g.k.}  \supset 
{{{\langle F_\Phi \rangle}_{ab}} \over {M_P}}
\lambda^a \lambda^b +h.c., 
\end{equation}
where $\lambda^{a,b}$ are gaugino fields. Here, we  denote 
by $\lambda^1$, $\lambda^2$ and $\lambda^3$  the 
$U(1)_{Y}$, $SU(2)_{L}$ and $SU(3)_{C}$ 
gaugino fields, respectively. Since the gauginos belong to the adjoint 
representation of the gauge group, 
$\Phi$ and $F_\Phi$ can belong to any of the 
representations appearing in the symmetric product of the 
two adjoint  representations of the unified gauge group. 

For example, in the case where the  SM gauge group 
is embedded  in the  grand unified gauge group $SU(5)$
the symmetric product of the two adjoint~({\bf 24} dimensional)
representations of $SU(5)$ leads to
\begin{equation}
({\bf 24 \otimes 24})_{Symm} = {\bf 1 \oplus 24 \oplus 75 \oplus 200}.
\label{product}
\end{equation}
\noindent
In Table~\ref{tab1} we have summarized  the ratios of  gaugino masses 
which result when $F_{\Phi}$ belongs to different representations of 
$SU(5)$ in the decomposition~(\ref{product}).


\begin{table}[t!]
\renewcommand{\arraystretch}{1.0}
\begin{center}
  \begin{tabular}{||c|ccc|ccc||}
    \hline 
    $SU(5)$ & $M_1^G$ & $M_2^G$ & $M_3^G$ & 
    $M_1^{EW}$ & $M_2^{EW}$ & $M_3^{EW}$
    \\ \hline 
    {\bf 1} & 1 & 1
    & 1 & 1 & 2 & 7.1 \\ 
    & & & & & & \\    
    {\bf 24} & 1 & 3 & -2 & 1 & 6 & -14.3 \\
     & & & & & & \\    
     {\bf 75} & 1 &-$\frac{3}{5}$ &-$\frac{1}{5}$ & 1 & -1.18 & -1.41 \\
      & & & & & & \\    
      {\bf 200} & 1 & $\frac{1}{5}$ &$\frac{1}{10}$ &1 & 0.4 & 0.71
    \\ \hline
  \end{tabular}
  \end{center}
  \caption{\label{tab1} Ratios of the gaugino masses at the GUT scale
    in the normalization ${M_1}(GUT)$ = 1, and at the electroweak
    scale in the normalization ${M_1}(EW)$ = 1 
    for $F$-terms in different representations of $SU(5)$.
    These results are obtained by using 1-loop renormalization
    group equations.}
\renewcommand{\arraystretch}{1.0}
\end{table}
\noindent


Similarly, nonuniversal gaugino masses can arise from the embedding of MSSM in
a grand unified theory based on $SO(10)$ and $E_6$. For these gauge groups we have the 
decomposition

$SO(10):$

\begin{equation}
({\bf 45 \otimes 45})_{Symm} = {\bf 1 \oplus 54 \oplus 210 \oplus 770};
\label{product2}
\end{equation}

$E_6:$

\begin{equation}
({\bf 78 \otimes 78})_{Symm} = {\bf 1 \oplus 650 \oplus 2430}.
\label{product3}
\end{equation}
\noindent
The analogs of Table~\ref{tab1} for the gauge groups SO(10) and $E_6$ are given in Appendix A of ~\cite{Ananthanarayan:2013fga},
where a detailed discussion has been presented on the 
phenomenological consequences of different choices of the grand unified 
gauge group.
We note that in the present work  we are not necessarily looking at very 
light neutralinos 
as the second lightest Higgs boson can be  heavy, and can decay into 
neutralinos that are massive. 

\subsection{Composition of Neutralinos}

The composition of the lightest neutralino $\tilde{\chi}_1^0$ 
in terms of the gauginos and Higgsinos can be 
written as~\cite{Gogoladze:2002xp,Bertone:2004pz}
\begin{equation}
 \tilde{\chi}_1^0 = Z_{11}\tilde{B}+Z_{12}\tilde{W}^3+
 Z_{13}\tilde{H}_1^0+Z_{14}\tilde{H}_2^0
\end{equation}
where
\begin{equation}
 Z_{1i}=\left(1,~-\frac{\cot \theta_W (M_1-m_{\chi^{0}_{1}})}{M_2-m_{\chi^{0}_{1}}},~\frac{(M_1-m_{\chi^{0}_{1}})(\mu \sin \beta + m_{\chi^{0}_{1}} \cos \beta)}{m_Z \sin \theta_W (m_{\chi^{0}_{1}} + \mu \sin 2\beta)}
,~-\frac{(M_1-m_{\chi^{0}_{1}})(\mu \cos \beta + m_{\chi^{0}_{1}} \sin \beta)}{m_Z \sin \theta_W (m_{\chi^{0}_{1}} + \mu \sin 2\beta)}\right  ).
\label{Z_{1i}}
\end{equation}
\noindent
The composition of the second lightest neutralino can be written as
\begin{equation}
 \tilde{\chi}_2^0 = Z_{21}\tilde{B}+Z_{22}\tilde{W}^3+
 Z_{23}\tilde{H}_1^0+Z_{24}\tilde{H}_2^0
\end{equation}
where
\begin{equation}
Z_{2i} =\left(1,~-\frac{\cot \theta_W (M_1-m_{\chi^{0}_{2}})}{M_2-m_{\chi^{0}_{2}}},~\frac{(M_1-m_{\chi^{0}_{2}})(\mu \sin \beta + m_{\chi^{0}_{2}} \cos \beta)}{m_Z \sin \theta_W (m_{\chi^{0}_{2}} + \mu \sin 2\beta)}
,~-\frac{(M_1-m_{\chi^{0}_{2}})(\mu \cos \beta + m_{\chi^{0}_{2}} \sin \beta)}{m_Z \sin \theta_W (m_{\chi^{0}_{2}} + \mu \sin 2\beta)}\right).
\label{Z_{2i}}
\end{equation}

\bigskip

\begin{figure}
\begin{minipage}[h]{0.45\linewidth}
\centering
\includegraphics[width=6cm, height=5cm]{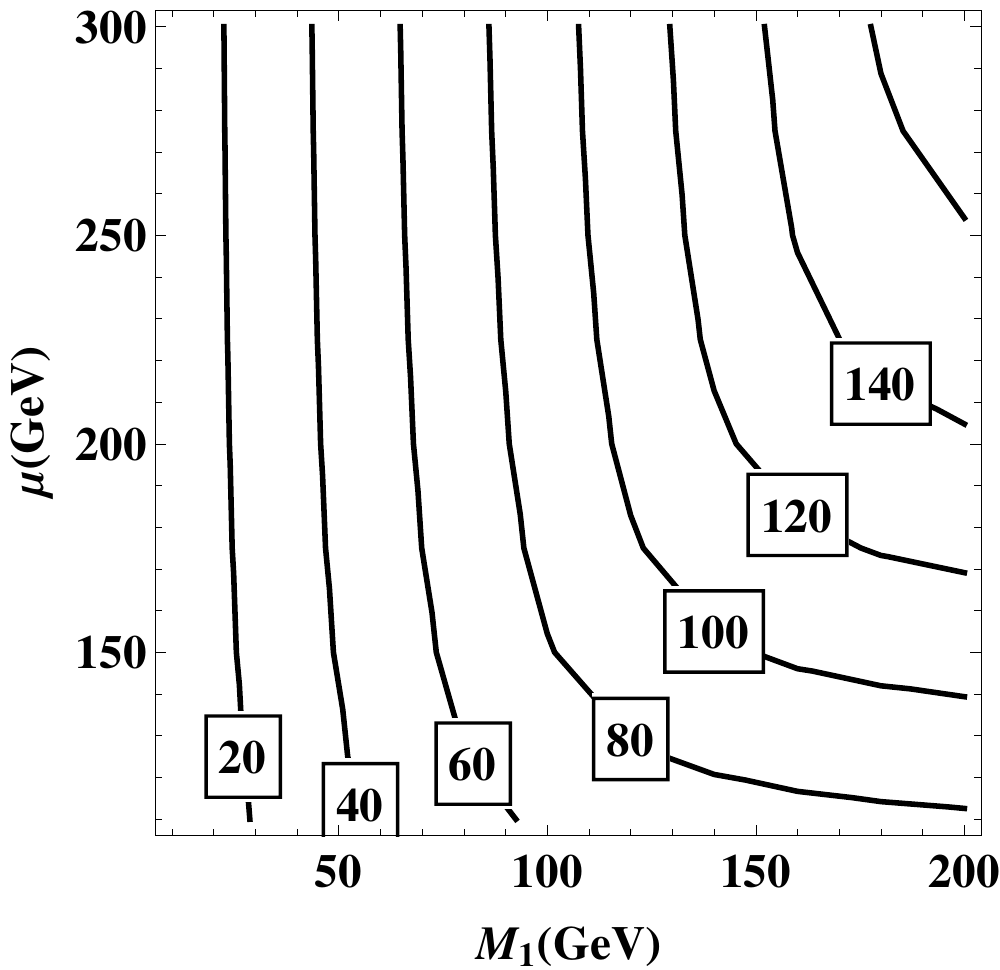}
\caption{The contours of constant lightest neutralino mass 
$m_{\tilde{\chi} ^{0}_{1}}$ in MSSM with arbitrary gaugino masses
at the GUT scale in the $\mu- M_{1}$ plane. Here the value of 
the parameter $M_{2} = 200$ GeV.}
\label{fig:M_chi1}
\end{minipage}
\hspace{0.4cm}
\begin{minipage}[h]{0.45\linewidth}
\centering
\includegraphics[width=6cm, height=5cm]{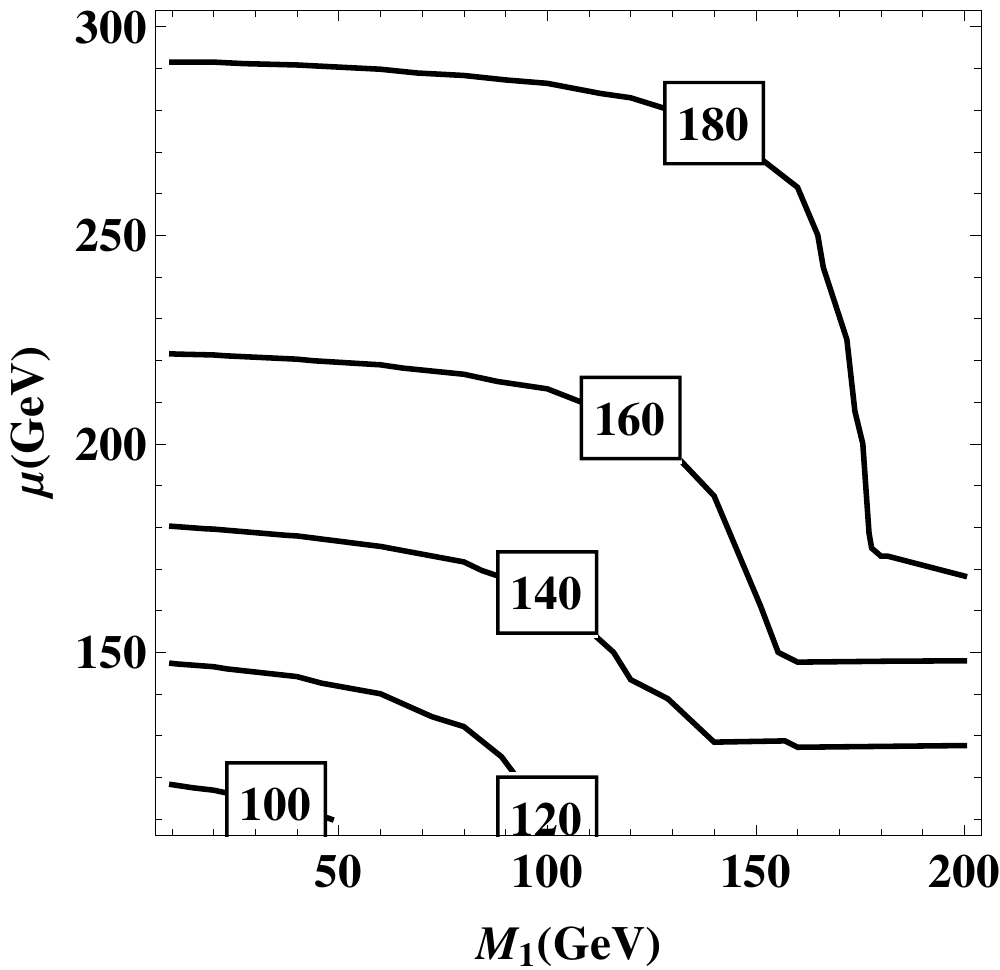}
\caption{The contours of constant second lightest neutralino mass 
$m_{\tilde{\chi} ^{0}_{2}}$ in MSSM with arbitrary gaugino masses 
at the GUT scale in the $\mu- M_{1}$ plane. Here the value
of $M_{2} = 200$ GeV.}
\label{fig:M_chi2}
\end{minipage}
\end{figure}
\begin{figure}
\begin{minipage}[h]{0.45\linewidth}
\centering
\includegraphics[width=6cm, height=5cm]{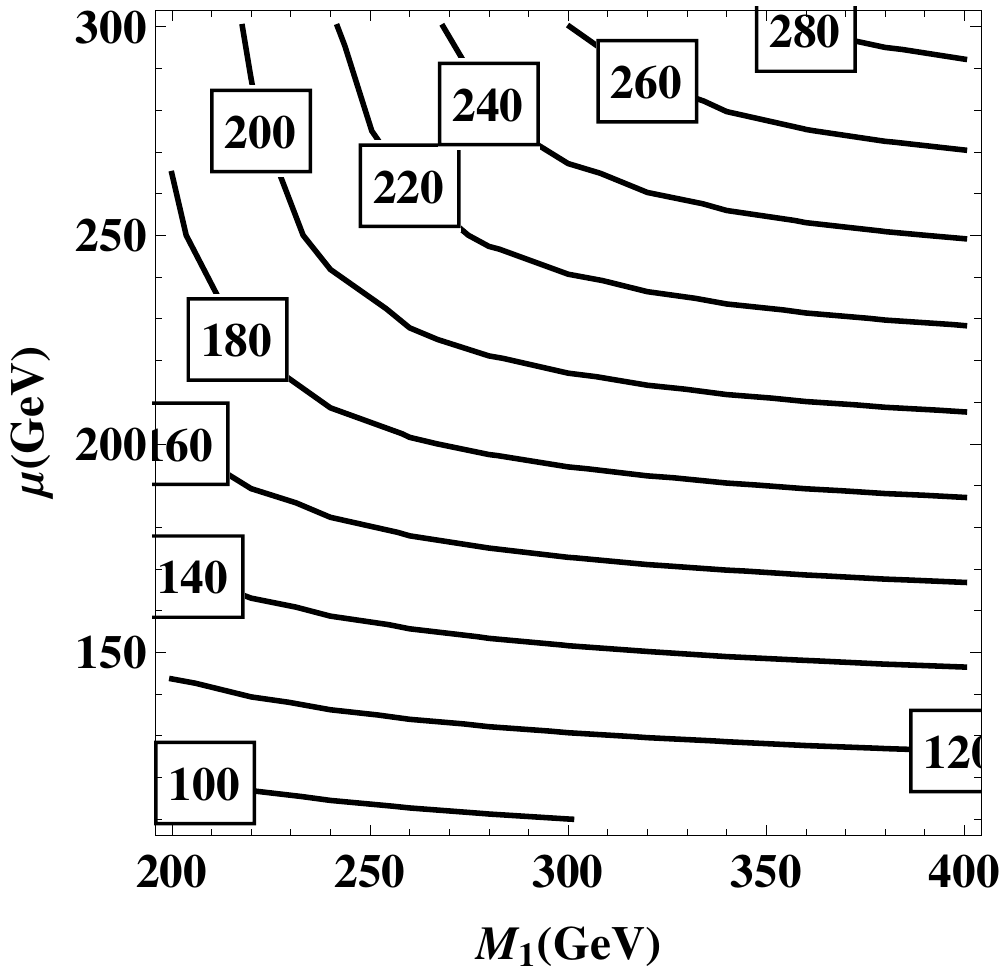}
\caption{The contours of constant lightest neutralino mass 
$m_{\tilde{\chi} ^{0}_{1}}$ in the $\mu- M_{1}$ plane
in MSSM with universal soft gaugino masses at the GUT scale.}
\label{fig:M_chi1_uni}
\end{minipage}
\hspace{0.4cm}
\begin{minipage}[h]{0.45\linewidth}
\centering
\includegraphics[width=6cm, height=5cm]{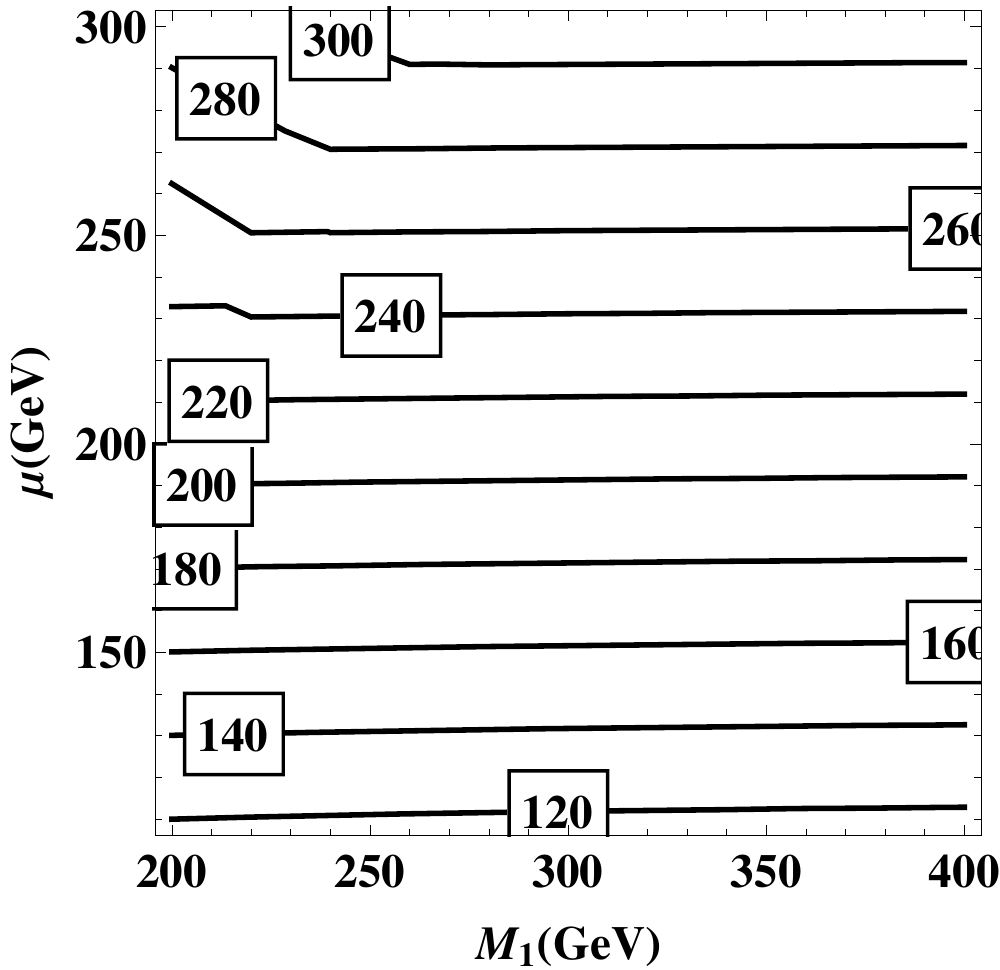}
\caption{The contours of constant second lightest neutralino mass 
$m_{\tilde{\chi} ^{0}_{2}}$ in the $\mu- M_{1}$ plane
in MSSM with universal soft gaugino masses at the GUT scale.}
\label{fig:M_chi2_uni}
\end{minipage}
\end{figure}

\begin{figure}
\begin{minipage}[h]{0.45\linewidth}
\centering
\includegraphics[width=6cm, height=5cm]{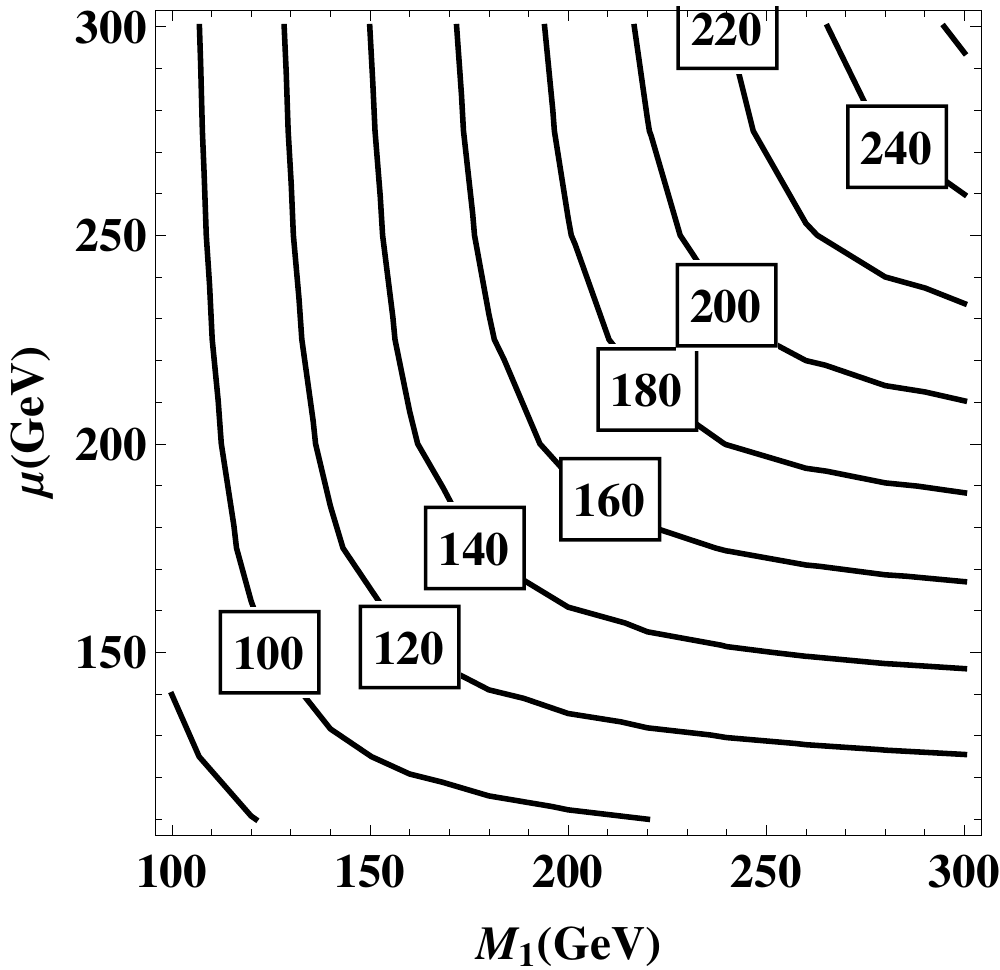}
\caption{The contours of constant lightest neutralino mass 
$m_{\tilde{\chi} ^{0}_{1}}$ in the $\mu- M_{1}$ plane
with MSSM embedded in  SO(10),  and with non-universal soft gaugino masses
at the GUT scale.}
\label{fig:M_chi1_nonuni1}
\end{minipage}
\hspace{0.4cm}
\begin{minipage}[h]{0.45\linewidth}
\centering
\includegraphics[width=6cm, height=5cm]{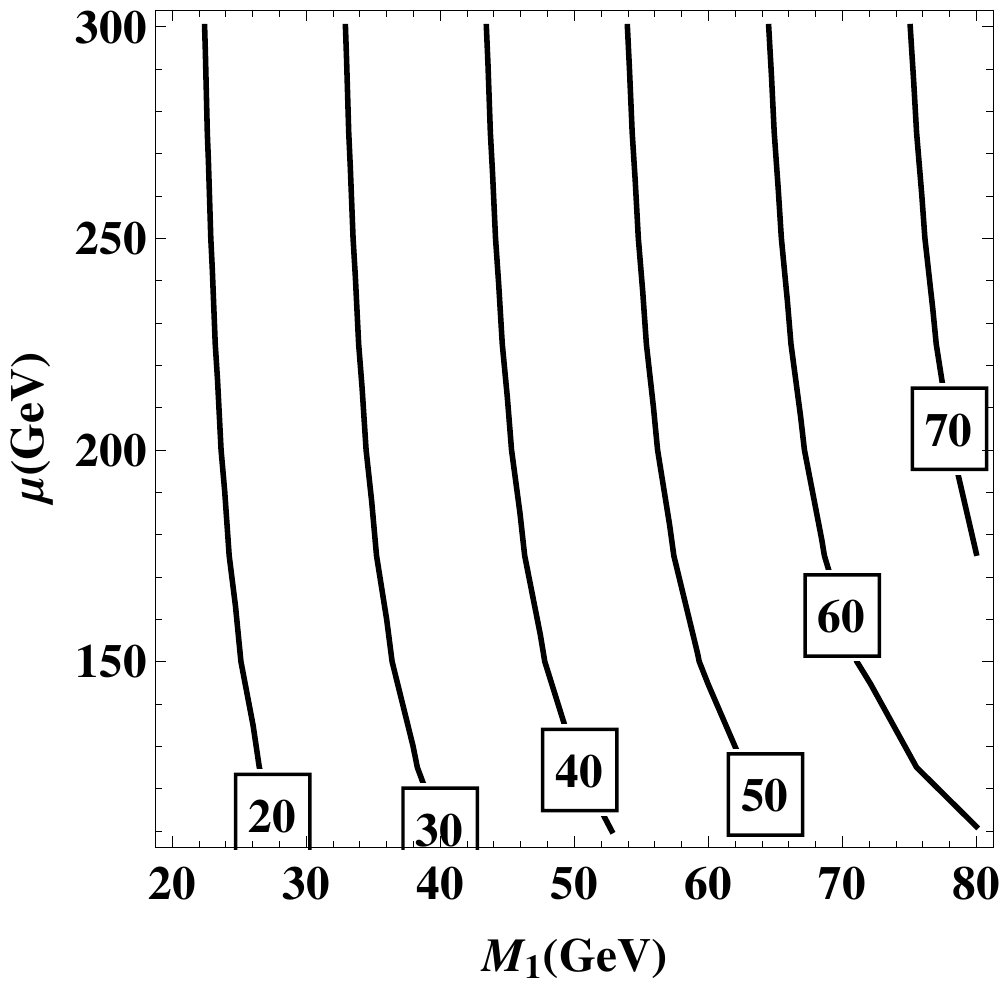}
\caption{The contours of constant second lightest neutralino mass 
$m_{\tilde{\chi} ^{0}_{1}}$ in the $\mu- M_{1}$ plane with
MSSM embedded in  $E_6$, and with 
non-universal soft gaugino masses at the GUT scale.}
\label{fig:M_chi1_nonuni2}
\end{minipage}
\end{figure}

  
\noindent
The possibility of $H^0$ decaying into neutralinos depends on the mass of 
the neutralinos into which it can decay. Our objective is to study 
the invisible decay of the second lightest Higgs boson into neutralinos.
For this purpose, we carry out a 
study of the  neutralino mass as a function of  $M_1$ and $\mu$, as these
parameters, apart from $M_2$, control the neutralino mass matrix, and hence
the neutralino mass eigenvalues.
We have considered  the constant mass curves for both the lightest and 
the second lightest neutralinos. 
Furthermore, we have considered each of the scenarios,
namely the case of arbitrary gaugino masses, universal,
and non-universal gaugino
masses, respectively at the grand unified scale in Figs.~\ref{fig:M_chi1},
~\ref{fig:M_chi2}, ~\ref{fig:M_chi1_uni}, 
\ref{fig:M_chi2_uni}, ~\ref{fig:M_chi1_nonuni1} 
and ~\ref{fig:M_chi1_nonuni2}.

For arbitrary supersymmetry breaking gaugino masses at the grand unified scale,
we shall choose the parameter  $M_2 = 200$ GeV, a choice we shall discuss
in the following. It is obvious that choosing universal or nonuniversal
boundary conditions at the grand unified scale is a rather restrictive 
condition. 

In Figs.~\ref{fig:M_chi1} and ~\ref{fig:M_chi2} we plot the contours 
of constant lightest and the second lightest neutralino masses
in the $\mu- M_1$ plane for arbitrary gaugino masses at the GUT scale. 
It is clear that $m_{\tilde{\chi} ^{0}_{1}}$ 
increases with $M_1$ but  does not change significantly with $\mu$. 
On the other hand the second lightest 
neutralino mass $m_{\tilde{\chi} ^{0}_{2}}$ does not change appreciably
with $M_1$ but  increases with $\mu$. This is primarily because 
$\tilde{\chi} ^{0}_{1}$ is bino-like, and $\tilde{\chi} ^{0}_{2}$ 
is a Higgsino-like state.
For low values of $M_1,$ the mass of $\tilde{\chi} ^{0}_{1}$ is almost
equal to $M_1,$ and it is bino-like for all values of $\mu.$
Furthermore, when $M_1 > \mu,$  $\tilde{\chi} ^{0}_{1}$ is 
Higgsino-like. At large $\mu,$ and when $M_1 > M_2 (200 \rm GeV),$
$\tilde{\chi} ^{0}_{1}$ will be wino-like.
We see from  Fig.~\ref{fig:M_chi1}  that with arbitrary gaugino masses 
and with $M_1$ larger than  150 GeV and $\mu$
larger than  250 GeV  the mass of  $\tilde{\chi} ^{0}_{1}$ approaches 
150 GeV and, thus, 
beyond these values a 300 GeV Higgs cannot decay into two lightest 
neutralinos. 

For low values of $M_1$ and with $\mu < M_2(200 \rm GeV)$, 
$\tilde{\chi} ^{0}_{2}$ is Higgsino-like. Furthermore, for 
values of $\mu > M_2 (200 \rm GeV)$, $\tilde{\chi} ^{0}_{2}$
is wino-like for all values of $M_1.$ For values of $\mu < M_1$,
$\tilde{\chi} ^{0}_{2}$ is Higgsino-like.
In Fig.~\ref{fig:M_chi2} we  see that for  $M_1 \geq 150$ GeV and
$\mu \geq 200$ GeV, the mass of  $\tilde{\chi} ^{0}_{2}$ approaches 
150 GeV, and hence for values larger than these  a 300 GeV
second lightest Higgs cannot decay into two second lightest neutralinos. 
Thus, only when  $M_1$ is large and $\mu$ is relatively small, or if
$\mu$ is large and $M_1$ is relatively small, $H^0$ can decay
invisibly into neutralinos.

\bigskip

It is instructive  to consider here the case of universal boundary 
conditions on the gaugino masses at the grand unified scale.
Considering the values of $M_1$ in the range $ 200 - 400 \rm GeV,$
which implies values of$M_2$ in the range of $400 - 800 \rm GeV,$
for low values of $\mu,$  ${\tilde{\chi} ^{0}_{1}}$ is Higgsino-like,
and for large values of $\mu,$ where $\mu > M_1, \mu,$  
$m_{\tilde{\chi} ^{0}_{1}}$ is bino-like. Similarly, for values of
$M_2 > M_1, \mu$  $m_{\tilde{\chi} ^{0}_{2}}$ is Higgsino-like.
In  Fig.~\ref{fig:M_chi1_uni} and Fig.~\ref{fig:M_chi2_uni} 
we plot the contours of constant neutralino masses 
in the $\mu- M_1$ plane for the case of universal boundary conditions on 
the soft gaugino masses at the grand unified scale. 
Here $m_{\tilde{\chi} ^{0}_{1}}$ and $m_{\tilde{\chi} ^{0}_{2}}$  do not change significantly 
as a function of $M_1$. However, both the masses are an increasing
function of $\mu$.
From Fig.~\ref{fig:M_chi1_uni} we see that with $\mu$ larger than 175 GeV,  
$m_{\tilde{\chi} ^{0}_{1}}$  
approaches 150 GeV, and, thus, a  $300$ GeV Higgs cannot 
decay into two lightest neutralinos for the case of universal boundary 
conditions. For  values of $\mu$  larger than 275 GeV 
$m_{\tilde{\chi} ^{0}_{1}}$ approaches
250 GeV for $M_1 \geq 300$ GeV. Hence, for these values of $\mu$, a
500 GeV Higgs cannot 
decay into lightest neutralinos..
However, for low $M_1$ larger values of  $\mu$  will be allowed. 
In Fig.~\ref{fig:M_chi2_uni} we can 
see $m_{\tilde{\chi} ^{0}_{2}}$ dominantly depends on $\mu$. 
For $\mu$ larger than 150 GeV,
$m_{\tilde{\chi} ^{0}_{2}}$ crosses the 150 GeV limit and the 
decay of the 300 GeV second lightest Higgs into two 
second lightest neutralinos will be forbidden for these values of $\mu$
for any value of  $M_1$. Similarly,  for a $500$ GeV 
Higgs boson, $\mu$ = 250 GeV is the limit because for higher values 
of $\mu$,   $m_{\tilde{\chi} ^{0}_{2}}$ approaches a value of $250$ GeV.

In Fig.~\ref{fig:M_chi1_nonuni1} and ~\ref{fig:M_chi1_nonuni2} we discuss 
two typical examples of  nonuniversal boundary conditions on the gaugino masses 
at the GUT scale. The first is the case of SO(10) GUT with 
the condition that the three gaugino masses are in the 
ratio $M_1:M_2:M_3$=$1:6:-14.3,$ and the second  one is with 
$E_6$ GUT with the boundary condition in which the three gaugino masses 
are in the ratio $M_1:M_2:M_3$ = $1:50.2:70.9$. 
The case of $SO(10)$ is akin to that universal boundary condtion
discussed above because the range of $M_2$ values considered is greater than
the range of $M_1$ and $\mu.$ The range of values of $M_1$ 
considered here is $ 100 - 300 \rm GeV.$ For $M_1 < \mu,$
${\tilde{\chi} ^{0}_{1}}$ is bino dominated, and for $M_1 > \mu,$
${\tilde{\chi} ^{0}_{1}}$ is Higgsino dominated.
On the other hand, the particular 
choice of $E_6$ representation is very interesting because 
this is the only GUT representation in which the lightest CP-even 
Higgs $h^0$ can decay into two lightest neutralinos. With 
$E_6$ boundary condition we can see that the mass of the lightest 
neutralinos does not  depend on $\mu,$  whereas in the case of SO(10) the 
mass depends on both $M_1$ and $\mu$. For this $E_6$ representation
the range of $M_1$ values considered here is  $ 20 - 80 \rm GeV,$
and values of $M_1 < \mu, M_2$ for the entire parameter space.
Hence in this case ${\tilde{\chi} ^{0}_{1}}$ is always bino-like.


\bigskip

We recall here that there are additional constraints coming
from the LEP analysis of the $Z^0$ decay in invisible modes, 
on the $\mu-M_1$ parameter
space. The invisible decay of the lightest Higgs boson to the lightest
neutralinos, if kinematically allowed,  is mainly constrained by the
$Z$ invisible decay rate.  This invisible decay width
has been measured very precisely by the LEP
experiments~\cite{ALEPH:2005ab} with  
\begin{equation}\label{z_width}
 \Gamma(Z^0\rightarrow \tilde{\chi}_1^0 \tilde{\chi}_1^0) < 3~{\rm MeV}.
\end{equation}
The $Z$ width to a pair of lightest neutralinos 
can be written as \cite{Heinemeyer:2007bw}
\begin{equation}
 \Gamma(Z^0\rightarrow \tilde{\chi}_1^0 \tilde{\chi}_1^0) =
 \frac{G_F m_Z^3}{6\sqrt{2}\pi}(Z_{13}^2-Z_{14}^2)
 \left(1-\frac{4m_{\tilde{\chi}_1^0}^2}{m_{Z^0}^{2}}\right)^{3/2}.
\end{equation}
\noindent
For our analysis we have taken $\tan \beta$ = 10~\cite{Heinemeyer:2012}.
The trilinear soft supersymmetry breaking coupling $A_t$ has been adjusted
in order to obtain a lightest Higgs boson  mass of $\approx$ 126 GeV. 
The gluino mass is taken to be 
1400 GeV~\cite{ATLAS:2012sma}, and the squarks are 
assumed to have a mass above 1 TeV~\cite{CMS:2012yua},
thereby respecting the current  experimental bounds.

In our calculations we have imposed the constraint of the 
lightest  chargino mass bound 
of $m_{\tilde{\chi}^+} > 94$ GeV following from the LEP experiments as well as 
the bound from invisible
$Z^0$ decay width coming from $Z^0$ decay into 
neutralinos~\cite{Dreiner:2012ex}.

Having discussed in some detail the correlations among the 
parameters of the neutralino sector, we 
now  turn to the implications of this analysis for  the possible
invisible decays of $H^0$ for  a  choice of parameters which are
consistent with the constraints discussed  in
this section.

\section{ Decay of the second lightest Higgs boson to neutralinos in the 
Minimal Supersymmetric Standard Model. }

We now address the main issue of this paper, namely the invisible decays 
of the heavier CP-even Higgs boson $H^0$ into neutralinos in the 
MSSM~\cite{Griest:1987qv,Gunion:1987fr}. We consider these decays in 
the decoupling regime of the MSSM, where the Higgs boson $H^0$ is 
relatively heavy. In the decoupling regime, the heavy $H^0$ can 
decay into neutralinos which are not necessarily light. 

   For this purpose we catalog the decay widths of the heavier CP-even 
Higgs boson into a pair of lighter neutralinos in the MSSM. 
These can be written as 

\begin{equation}\label{higgs_decay1}
\Gamma (H^0 \rightarrow \tilde{\chi}_1^0 \tilde{\chi}_1^0) 
= \frac{G_F M_W^2 M_H}{2 \sqrt{2} \pi}(1- 4m^2_{\tilde{\chi}^0_1}/M_H^2)^{3/2}
\left[(Z_{12} - \tan{\theta_W} Z_{11})
(Z_{13}\cos \alpha - Z_{14}\sin \alpha)\right]^2,
\end{equation}
\noindent
where $Z_{ij}$ are the elements of the matrix $Z$ which diagonalizes the 
neutralino mass matrix (\ref{matrix1}). In the decoupling limit, 
when the mass of the pseudoscalar $A^0$ is large compared to the mass 
of the $Z$ boson $M_Z$, the Higgs mixing angle  
$\alpha \rightarrow \beta - \pi/2$, so that the decay 
width (\ref{higgs_decay1}) can be written as 

\begin{equation}\label{higgs_decay1_1}
\Gamma (H^0 \rightarrow \tilde{\chi}_1^0 \tilde{\chi}_1^0) 
= \frac{G_F M_W^2 M_H}{2 \sqrt{2} \pi}(1- 4m^2_{\tilde{\chi}^0_1}/M_H^2)^{3/2}
\left[(Z_{12} - \tan{\theta_W} Z_{11})
(Z_{13}\sin \beta + Z_{14}\cos \beta)\right]^2.
\end{equation}

\medskip

\noindent 

Similarly, the decay width of $H^0$ into a pair of 
second lightest neutralinos can be written as 

\begin{equation}
\label{higgs_decay2}
\Gamma (H^0 \rightarrow \tilde{\chi}_2^0 \tilde{\chi}_2^0)
= \frac{G_F M_W^2 M_H}{2 \sqrt{2} \pi}(1- 4m^2_{\tilde{\chi}^0_2}/M_H^2)^{3/2}
\left[(Z_{22} - \tan{\theta_W} Z_{21})
(Z_{23}\cos \alpha - Z_{24}\sin \alpha)\right]^2,
\end{equation}
which in the decoupling regime becomes
\begin{equation}
\Gamma (H^0 \rightarrow \tilde{\chi}_2^0 \tilde{\chi}_2^0)
= \frac{G_F M_W^2 M_H}{2 \sqrt{2} \pi}(1- 4m^2_{\tilde{\chi}^0_2}/M_H^2)^{3/2}
\left[(Z_{22} - \tan{\theta_W} Z_{21})
(Z_{23}\sin \beta + Z_{24}\cos \beta)\right]^2.
\end{equation}
On the other hand the decay width for the process $H^0 \rightarrow \chi_1^0
\chi_2^0$ can be written as
\begin{eqnarray}
\label{higgs_decay12}
\Gamma (H^0 \rightarrow \tilde{\chi}_1^0 \tilde{\chi}_2^0)
& = & \frac{G_F M_W^2 M_H}{\sqrt{2} \pi} F_{121}^2
[1 + (m^2_{\tilde{\chi}^0_1}/M_H^2 - m^2_{\tilde{\chi}^0_2}/M_H^2)^2
 - 2 (m^2_{\tilde{\chi}^0_1}/M_H^2 + m^2_{\tilde{\chi}^0_2}/M_H^2)]^{1/2} \\
\nonumber
&&\times [1 - m^2_{\tilde{\chi}^0_1}/M_H^2 -m^2_{\tilde{\chi}^0_2}/M_H^2
- 2(\epsilon_1 \epsilon_2 /M_H^2) m_{\tilde{\chi}^0_1}  m_{\tilde{\chi}^0_2}],
\end{eqnarray}
where
\begin{equation}
F_{121} = \frac{1}{2} (Z_{22} - \tan\theta_W Z_{21})(Z_{13} \cos\alpha -
Z_{14} \sin\alpha)
+ \frac{1}{2} (Z_{12} - \tan\theta_W Z_{11})(Z_{23} \cos\alpha -
Z_{24} \sin\alpha),
\end{equation}
which in the decoupling regime reduces to
\begin{equation}
F_{121} = \frac{1}{2} (Z_{22} - \tan\theta_W Z_{21})( Z_{13} \sin\beta +
Z_{14} \cos\beta)
+ \frac{1}{2} (Z_{12} - \tan\theta_W Z_{11})( Z_{23} \sin\beta +
Z_{24} \cos\beta), 
\label{F_{121}}
\end{equation}
and where the constants  $\epsilon_i$s (i=1,2,3,4) stand for the sign of 
the neutralino
mass. When the neutralino mass matrix is diagonalized,
we allow the sign of the ith eigenvalue to be  either positive or negative.

In our previous work ~\cite{Ananthanarayan:2013fga}, we 
had considered the possibility of the lightest Higgs boson decaying into 
lightest neutralinos in the MSSM and the NMSSM. 
As the mass of the lightest Higgs boson is fixed to be 126 GeV, 
we had considered the case of the lightest neutralino with mass 
$ m_{\tilde{\chi}^0_1}<m_{h^{0}/2}$. We had concluded that such a 
constraint on the neutralino mass is not satisfied with either 
universal or nonuniversal boundary conditions on the soft gaugino 
masses at the GUT scale. The only exception to this conclusion 
was a nonuniversal scenario with a very large dimensional 
representation$(\bf{2430})$ of the gauge group $E_6$. 
However, this is not an appealing possibility. The constraint on the 
light neutralino masses was obeyed in the MSSM only with arbitrary 
gaugino masses at the GUT scale. Since in the present paper we are 
considering the invisible decay of the heavier Higgs boson $H^0$, 
the lightest neutralinos need not be massless or very light. 
Thus, in the following we shall consider the case of massive neutralinos, 
which can arise in all the three cases, i.e. universal, nonuniversal and 
arbitrary soft gaugino masses at the GUT scale. 


\section{Results for the Invisible Decay of the Heavier Higgs boson}

\bigskip

Having summarized the results for the decay widths for the invisible 
decay of the heavier Higgs boson in the previous section, we now 
evaluate these decay widths using the parameter space of the MSSM 
allowed by the present experimental constraints. For this purpose, 
we shall use the boundary conditions as implied by the embedding of 
MSSM in a grand unified theory, as well as for arbitrary soft 
gaugino masses at the GUT scale. As discussed in the last section, 
the final states in the decay process that are of interest to us 
are $\chi_i^0 \chi_j^0, i=1,2$. 

We first summarize the parameter space used in our analysis. 
The trilinear soft supersymmetry breaking parameter $A_t$ pertaining 
to the stop  is adjusted to obtain the lightest Higgs boson mass in the 
range 123-127 GeV, although it does not affect the neutralino sector. 
To get the lightest Higgs mass in this range we use the "maximal mixing" 
scenario~\cite{Carena:2002qv}, wherein 
\begin{equation}
A_{t}- \mu \cot \beta = \sqrt{6} M_{S},
\end{equation}
\noindent
where $M_{S}$ is the soft supersymmetry-breaking scale, which we take 
to be 1.5 TeV. The gluino mass is taken to be 1.4 TeV, 
and squarks are assumed to have masses above 1 TeV. The Slepton masses are assumed to be greater than 500 GeV, 
thereby respecting all current experimental bounds~\cite{Lepton:2013}. 
For the case of arbitrary boundary condition on the gaugino mass parameters, 
we have taken $M_2=200$ GeV. Furthermore, we have also imposed the 
constraint of the lightest chargino mass $m_{{\tilde{\chi}}^{+}} >$  94 GeV 
as implied by the LEP experiments,  as well as the bound on the 
invisible $Z^0$ width into neutralinos. We note that decreasing the 
value of $M_2$ increases the chargino mass bound on $\mu$, 
thereby eliminating the parameter space with large invisible 
branching ratio for the lightest Higgs boson~\cite{Dreiner:2012ex}.

In addition, we have also imposed the constraints resulting from 
(g-2) of the muon, as well as flavor constraints following from 
$b \rightarrow s \gamma$ and $B_s \rightarrow \mu^{+} \mu^{-}$. 
We have implemented these constraints using CalcHEP. 
The input parameters so obtained are summarized 
in Table~\ref{tab:mssmparam1}~\cite{CMS:2014qv,Arbey:2012}

\begin{table}[htb]
\renewcommand{\arraystretch}{1.0}
\begin{center}
\vspace{0.5cm}
\begin{tabular}{|c|c|c|c|}
\hline
$\tan\beta$ = 10 &$M_S$ = 1.5 TeV &$M_A$ = 300,500 GeV 
&$M_2$ = 200 GeV \\
\hline
$M_3$ = 1402 GeV    &$A_t$= 3600 GeV &$A_b$= 3600 GeV &$A_{\tau}$= 1000 GeV \\
\hline 
\end{tabular}
\end{center}
\vspace{-0.5cm}
\caption{Input parameters for the  MSSM consistent with all the 
experimental constraints.}
\renewcommand{\arraystretch}{1.0}
\label{tab:mssmparam1}
\end{table}
\noindent
\bigskip
\begin{figure}
\begin{minipage}[h]{0.45\linewidth}
\centering
\includegraphics[width=6cm, height=5cm]{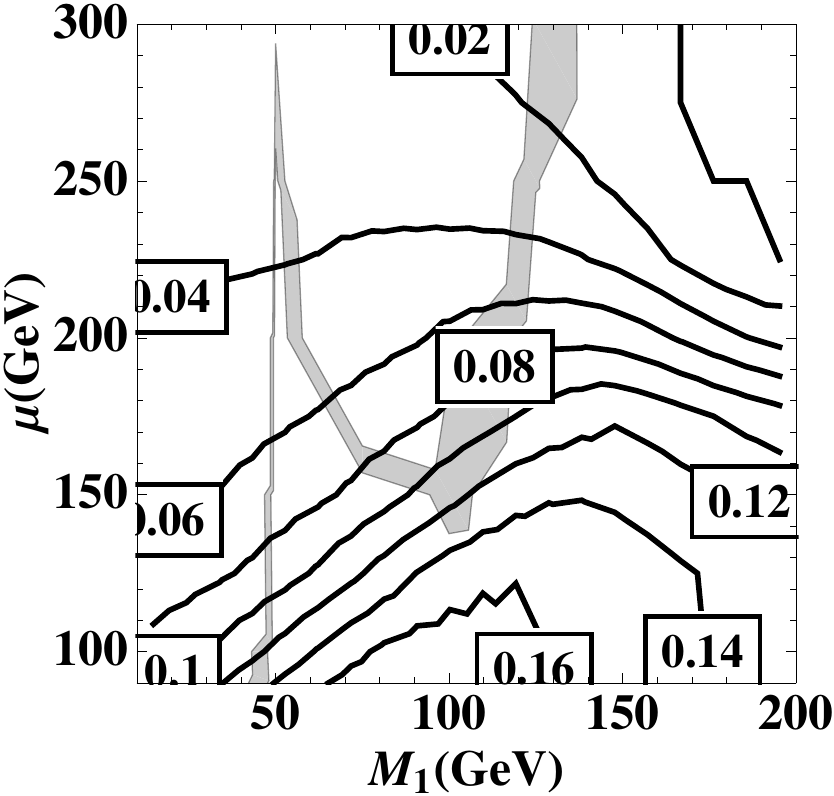}
\caption{The contours of constant branching ratio for  
($ H^0 \rightarrow \tilde{\chi} ^{0}_{1} \tilde{\chi} ^{0}_{1}$) 
in the $\mu- M_{1}$ plane for  $M_A = 300$ GeV with arbitrary gaugino masses  
at the GUT scale. Here  $M_{2}$ is taken to be 200 GeV. The shaded region
represents the region allowed by the WMAP data.}
\label{fig:xi1xi1_300}
\end{minipage}
\hspace{0.4cm}
\begin{minipage}[h]{0.45\linewidth}
\centering
\includegraphics[width=6cm, height=5cm]{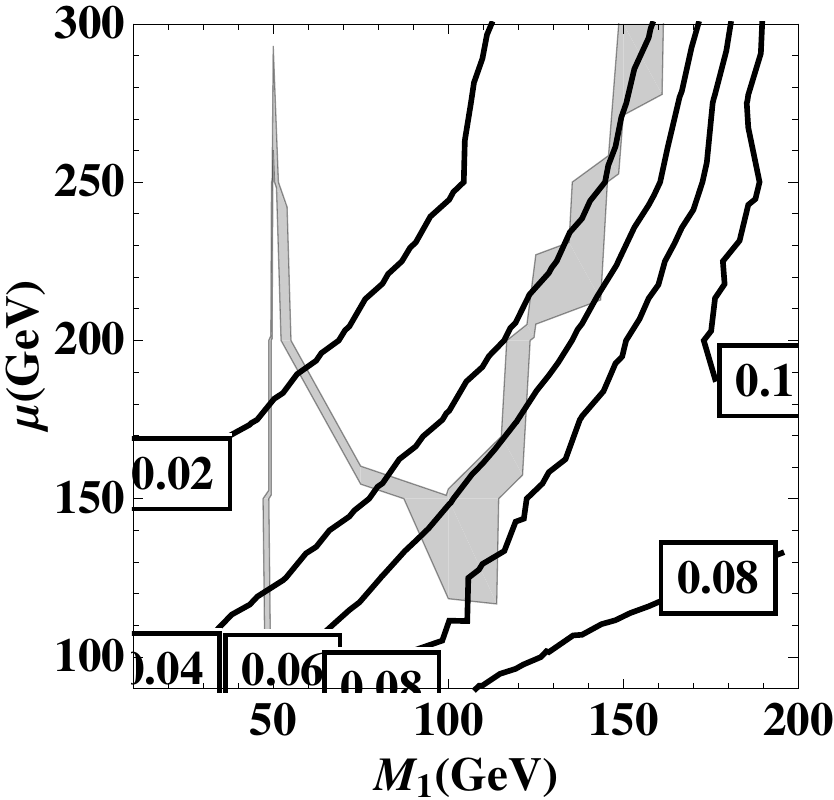}
\caption{The contours of constant branching ratio for  
($H^0 \rightarrow \tilde{\chi} ^{0}_{1} \tilde{\chi} ^{0}_{1}$) 
in the $\mu- M_{1}$ plane for $M_A = 500$ GeV with arbitrary gaugino masses 
at the GUT scale. Here $M_{2}$ is taken to be 200 GeV. The shaded region
is allowed by the WMAP data.}
\label{fig:xi1xi1_500}
\end{minipage}
\end{figure}  

\bigskip

We now proceed to present the results of our calculations. The results 
are presented as constant branching ratio contours in the 
$\mu-M_1$ plane for two different values of $M_A \approx M_H = 300$ GeV, 
$500$ GeV. For the case of arbitrary soft supersymmetry breaking gaugino mass 
parameters we have taken $M_2=200$ GeV, and $\tan \beta = 10.$ 
In Fig.~\ref{fig:xi1xi1_300} we observe that the branching ratio for  
$ H^0 \rightarrow \tilde{\chi} ^{0}_{1} \tilde{\chi} ^{0}_{1}$ can be at 
most 16\% for $M_A=300$ GeV, and that too in a very narrow region of 
the parameter space. Furthermore, from  Fig.~\ref{fig:xi1xi1_500} 
we can see that this branching fraction reduces with the increase in 
the value of $M_A$, having a value of 10\% for $M_A=500$GeV.

Since we are considering MSSM with $R_p$ conservation, the lightest 
neutralino is an absolutely stable particle. In this case it is 
important to check that  it  is not overproduced
in the early universe. We have, therefore, 
calculated the relic density of the lightest neutralino for the range
of parameters used in our calculations using 
micrOMEGAs~\cite{ Belanger:2014vza},
and imposed the constraints  following from the  
WMAP data~\cite{Bennett2003, Spergel2003}. The result of this calculation
is shown  in Fig.\ref{fig:xi1xi1_300} as the shaded region allowed
by the constraints of the relic density considerations. Similarly, in
Fig.\ref{fig:xi1xi1_500} the shaded region is the region allowed by the
WMAP data.

\medskip

This behavior of the decay  
$H^0 \rightarrow \tilde{\chi} ^{0}_{1} \tilde{\chi} ^{0}_{1}$ 
may be understood as follows. At low values of $M_1$,
$\chi_i^0$  is bino-like for all values of $\mu$
in the region of the parameter space that we are considering.
The Higgs decay to neutralino is suppressed because of small 
value of $Z_{14}$ for $\tan \beta=10$. 
The mass of the lightest neutralino increases with the increase 
in the value of $M_1$. In Fig.~\ref{fig:xi1xi1_300}, for $M_1 \geq 150$ 
GeV there is a kinematic suppression for $M_A=300$ GeV, resulting in a 
smaller branching ratio. On the other hand for $M_A=500$ GeV there is 
no kinematic suppression, hence the branching ratio does not fall off, 
as is clear from Fig.~\ref{fig:xi1xi1_500}.

\medskip

For small values of $M_1$ and for all values of $\mu$ in the parameter
space that we are considering, the quantity 
$( M_1-m_{\tilde{\chi} ^{0}_{1}})$ 
is close to zero, so that the decay width for 
$H \rightarrow \tilde{\chi} ^{0}_{1} \tilde{\chi} ^{0}_{1}$ 
is reduced as can be seen from the analytical results in the 
previous Section, Eq.~(\ref {higgs_decay2}).
From Eq.~\ref{Z_{1i}}, 
we observe that $Z_{13}$ is almost zero. Hence the branching 
fraction for both values of $M_A=300$ GeV and $500$ GeV is small. 
For fixed $\mu$, with increasing $M_1 > \mu$, ${\tilde{\chi} ^{0}_{1}}$
is Higgsino-like, $ ( M_1-m_{\tilde{\chi} ^{0}_{1}})$ 
increases, the total width decreases, and hence the branching ratio increases. 
For fixed $M_1$ with decreasing $\mu,$  $( M_1-m_{\tilde{\chi} ^{0}_{1}})$ 
increases. Furthermore, the denominator(total width) decreases with 
increasing $\mu$, the numerator decreases faster, and hence the branching 
fraction decreases with $\mu$. Hence, to obtain a constant branching 
fraction, we have to increase $M_1$ and $\mu$ simultaneously 
until the factor $ (1- 4m^2_{\tilde{\chi}^0_1}/M_H^2)^{3/2}$ 
starts dominating, and causes kinematic suppression for the case 
of $M_A=300$ GeV.

We note that the total width has been calculated using CalcHEP, 
and depending on the values of parameters $M_1$and $\mu$ different
channels contribute to the total width, which has been used in the 
calculations presented here.

\bigskip

\begin{figure}
\begin{minipage}[h]{0.45\linewidth}
\centering
\includegraphics[width=6cm, height=5cm]{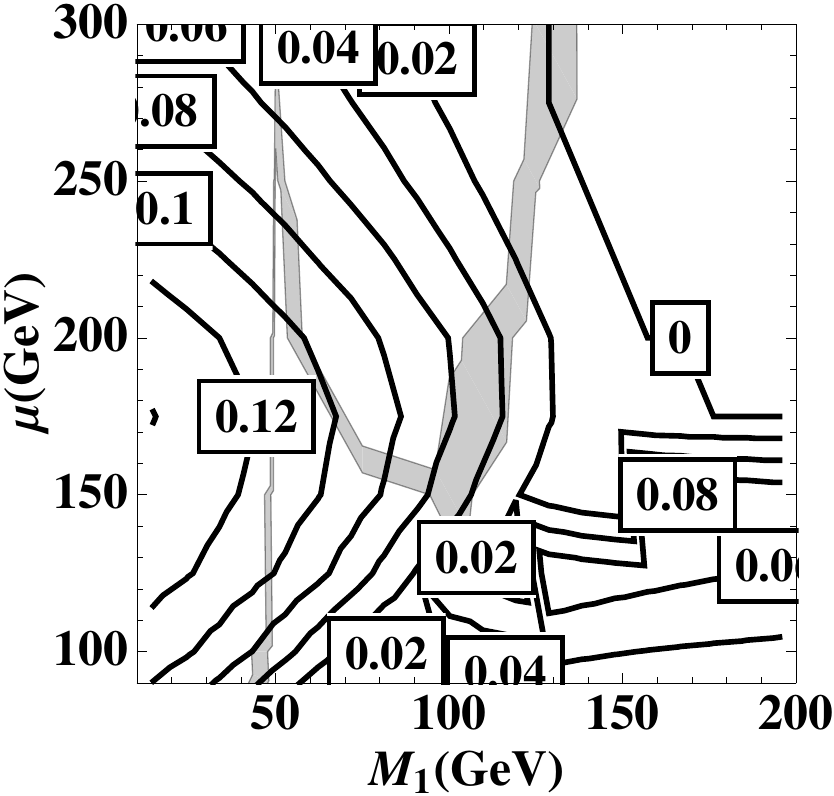}
\caption{The contours of constant branching ratio of 
($ H^0 \rightarrow \tilde{\chi} ^{0}_{1} \tilde{\chi} ^{0}_{2}$) 
in the $\mu- M_{1}$ plane for  $M_A = 300$ GeV
with  arbitrary gaugino masses at the GUT scale. 
Here $M_{2}$ is taken to be 200 GeV. The shaded region is allowed
by WMAP data.}
\label{fig:xi1xi2_300}
\end{minipage}
\hspace{0.4cm}
\begin{minipage}[h]{0.45\linewidth}
\centering
\includegraphics[width=6cm, height=5cm]{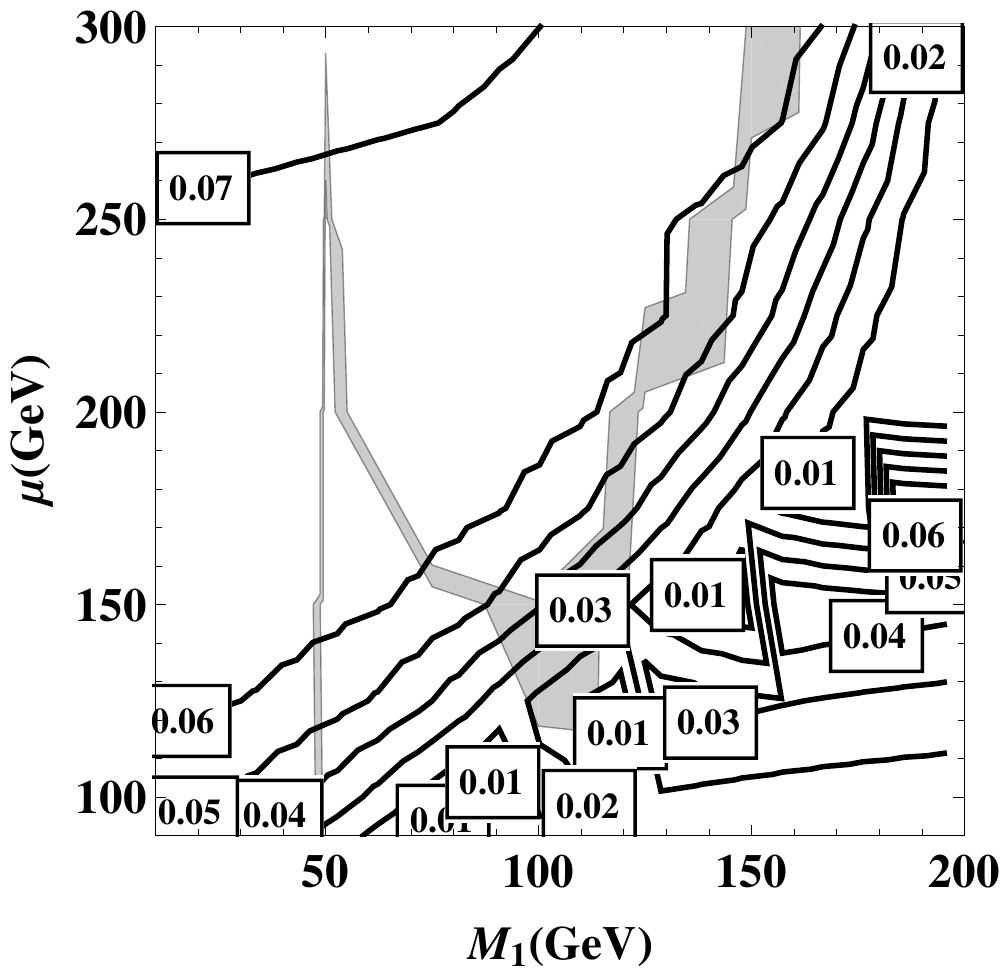}
\caption{The contours of constant branching ratio for  
($ H^0 \rightarrow \tilde{\chi} ^{0}_{1} \tilde{\chi} ^{0}_{2}$) 
in the $\mu- M_{1}$ plane for $M_A = 500$ GeV 
with arbitrary gaugino masses at the GUT scale. Here
$M_{2}$ is taken to be 200 GeV. The shaded region is allowed by 
WMAP data.}
\label{fig:xi1xi2_500}
\end{minipage}
\end{figure}

We may mention here that if we increase the value of $M_2 > 200 \rm GeV,$ 
large branching fractions for the invisible decay would still be possible. 
On the other hand,  values of $M_2 < 200 \rm GeV$ will exclude 
the region of parameter space with considerable invisible branching fraction.

\bigskip

We now turn to the invisible decay 
$ H^0 \rightarrow \tilde{\chi} ^{0}_{1} \tilde{\chi} ^{0}_{2}$ 
for values of $M_A \approx M_H=300$ GeV and $500$ GeV. 
In Fig.~\ref{fig:xi1xi2_300} and Fig.~\ref{fig:xi1xi2_500} 
we show the contours of constant branching ratios for this decay.  
For $M_A= 300$~GeV the branching ratio can at most be 12\% and for 
$M_A= 500$~GeV it is at most be 6\%. 
As we have seen in case of  the decay 
$H^{0} \rightarrow \chi^0_1 \chi^0_1$  
here also we see a kinematic suppression in case of 300 GeV Higgs boson 
for large values of $M_1$ and $\mu$. The reason is that the sum of the 
masses of the lightest and the second lightest neutralino reaches 
its limiting value and  the decay $H^{0} \rightarrow \chi^0_1 \chi^0_2$ 
is no longer possible. On the contrary, for 500 GeV Higgs boson there is 
no kinematic suppression, because $M_H$ which is nearly  equal to $M_A$ in 
the decoupling regime is sufficient to produce a lightest and a second
lightest neutralino for this particular region of the parameter space 
and the decay width is governed by the function $F_{121}$ 
in Eq.(~\ref{F_{121}}). We note that the shaded region in these
Figs. represents the region allowed by the WMAP data.

\bigskip

\begin{figure}
\begin{minipage}[h]{0.45\linewidth}
\centering
\includegraphics[width=6cm, height=5cm]{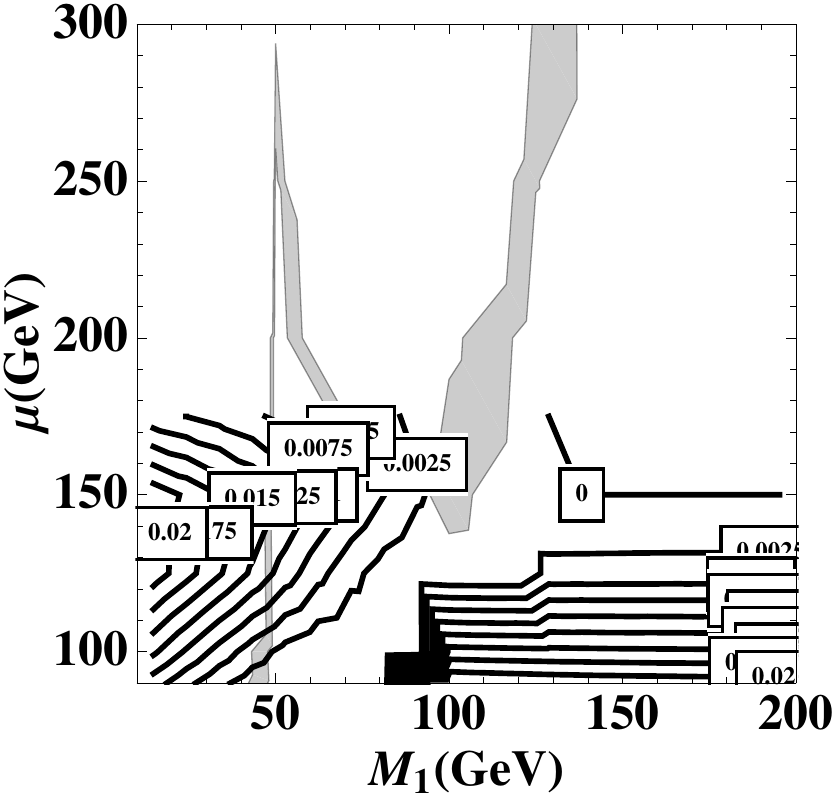}
\caption{The contours of constant branching ratio for 
($ H^0 \rightarrow \tilde{\chi} ^{0}_{2} \tilde{\chi} ^{0}_{2}$) 
in the $\mu- M_{1}$ plane for $M_A = 300$ GeV with 
arbitrary gaugino masses at the GUT scale. Here
$M_{2}$ is taken to be 200 GeV. The shaded region represents the region
allowed by the WMAP data.}
\label{fig:xi2xi2_300}
\end{minipage}
\hspace{0.4cm}
\begin{minipage}[h]{0.45\linewidth}
\centering
\includegraphics[width=6cm, height=5cm]{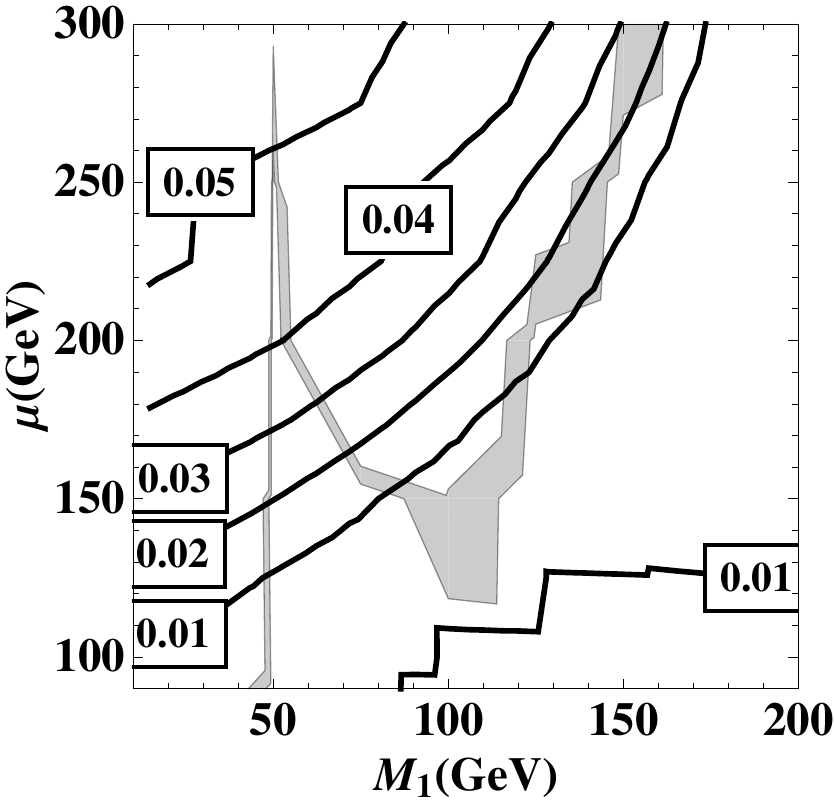}
\caption{The contours of constant branching ratio for
($H^0 \rightarrow \tilde{\chi} ^{0}_{2} \tilde{\chi} ^{0}_{2}$) 
in the $\mu- M_{1}$ plane for  $M_A = 500$ GeV with arbitrary 
gaugino masses in the $\mu- M_{1}$ plane. 
Here $M_{2}$ is taken to be 200 GeV. The shaded region is allowed 
by WMAP data.}
\label{fig:xi2xi2_500}
\end{minipage}
\end{figure}  

\bigskip

In Fig.~\ref{fig:xi2xi2_300} and Fig.~\ref{fig:xi2xi2_500} 
we show the contours of constant branching ratio 
for the decay 
$ H^0 \rightarrow \tilde{\chi} ^{0}_{2} \tilde{\chi} ^{0}_{2}$  
in the $\mu - M_{1}$ plane for values of 
$M_A \approx M_H = 300$ and $500$~GeV, respectively. 
The branching ratio can at most be 2\% in the case of 300 GeV 
second lightest Higgs, and it reaches a value of 
about 5\% in case of 500 GeV second lightest Higgs boson. 
There is a kinematic suppression for 300 GeV Higgs for values of 
$\mu$ larger than $200$~GeV as discussed earlier
but for 500 GeV Higgs there is no such suppression and 
the branching ratio actually increases with increasing $M_A$
in some regions of the  parameter space. The shaded region in these
Figs. represents the region allowed by the WMAP data.
 
The mass of the second lightest neutralino becomes almost equal 
to $M_1$ for large values of $\mu$ and $M_1$. Hence,
in this region the branching ratio goes to zero. 
And from the Fig.~\ref{fig:M_chi2} we can see that for values larger than
$M_1=100$ GeV, $m_{\tilde{\chi} ^{0}_{2}}$ does not depend on $M_1$ 
and depends only on $\mu$. Hence,
for a fixed values of $\mu$, the quantities  $Z_{22}$,
$Z_{23}$ and $Z_{24}$ increase with $M_1$ as $(M_1- m_{\chi_1^0})$ 
increases with $M_1$. Therefore, 
the width $H^0 \rightarrow \tilde{\chi} ^{0}_{2} \tilde{\chi} ^{0}_{2}$ 
also increases. We have checked that the total width decreases with $M_1$. 
So the branching fraction 
$H^0 \rightarrow \tilde{\chi} ^{0}_{2} \tilde{\chi} ^{0}_{2}$ 
increases with $M_1$. Now for a fixed $M_1,$ 
the quantity $(M_1- m_{\chi_1^0})$ decreases with $\mu$. Furthermore,
$Z_{22}$ decreases with $\mu$. $Z_{23}$ and $Z_{24}$ also decrease, having $(M_1- m_{\chi_1^0})$ 
in the numerator and $(m_{\chi_0^2} + \mu \sin 2 \beta)$ in the denominator. 
Also, the total width decreases with $\mu$, but the numerator 
in the branching ratio decreases much faster. Hence the branching 
fraction $H^0 \rightarrow \tilde{\chi} ^{0}_{2} \tilde{\chi} ^{0}_{2}$ 
decreases with $\mu$. Let us recall the branching ratio increases with $M_1$. 
Consequently we have to increase both $M_1$ and $\mu$ 
to get the contours of constant branching ratio. This is reflected 
in  Figs.~\ref{fig:xi2xi2_300} and ~\ref{fig:xi2xi2_500}.

\bigskip

\begin{figure}
\begin{minipage}[b]{0.45\linewidth}
\centering
\includegraphics[width=6cm, height=5cm]
{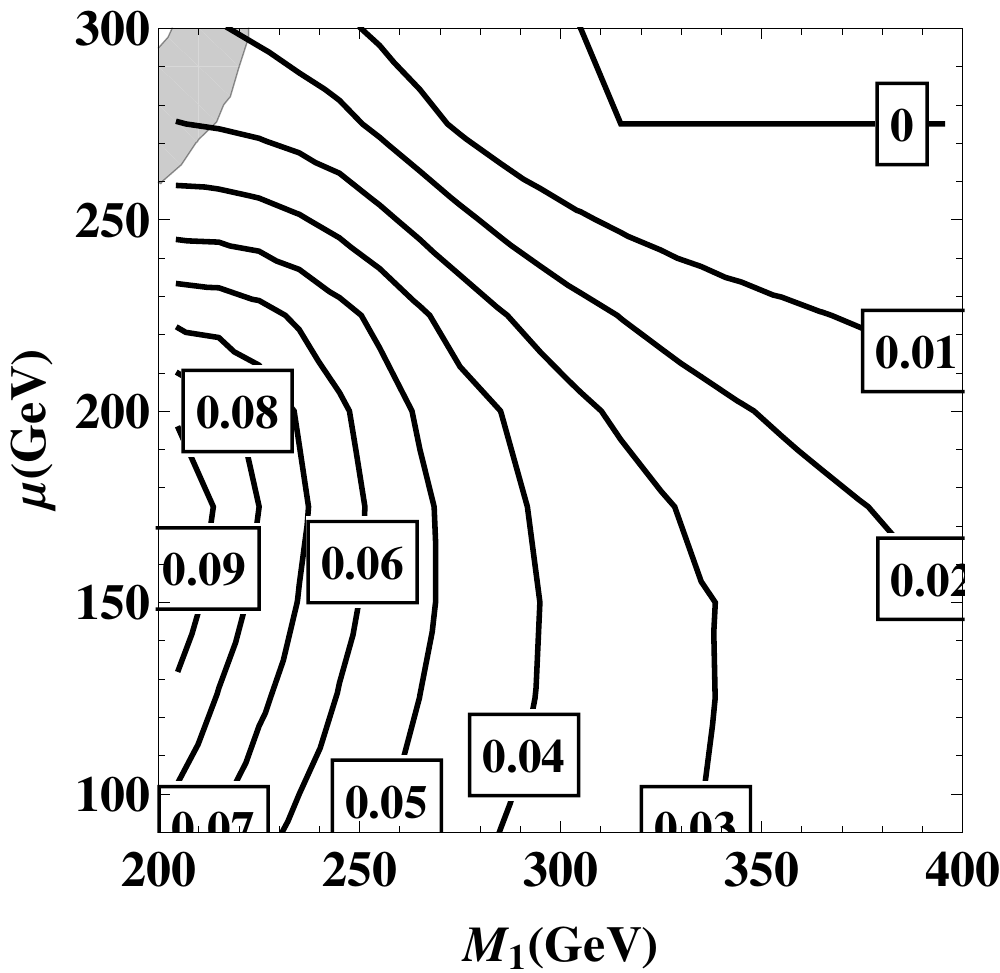}
\caption{The contours of constant branching ratio for
($ H^0 \rightarrow \tilde{\chi} ^{0}_{1} \tilde{\chi} ^{0}_{1}$) 
in the $\mu- M_{1}$ plane for  $M_A = 500$ GeV with universal 
gaugino masses at the GUT scale. The shaded region is allowed by 
WMAP data.}
\label{fig:xi1xi1_500_uni}
\end{minipage}
\hspace{0.4cm}
\begin{minipage}[b]{0.45\linewidth}
\centering
\includegraphics[width=6cm, height=5cm]
{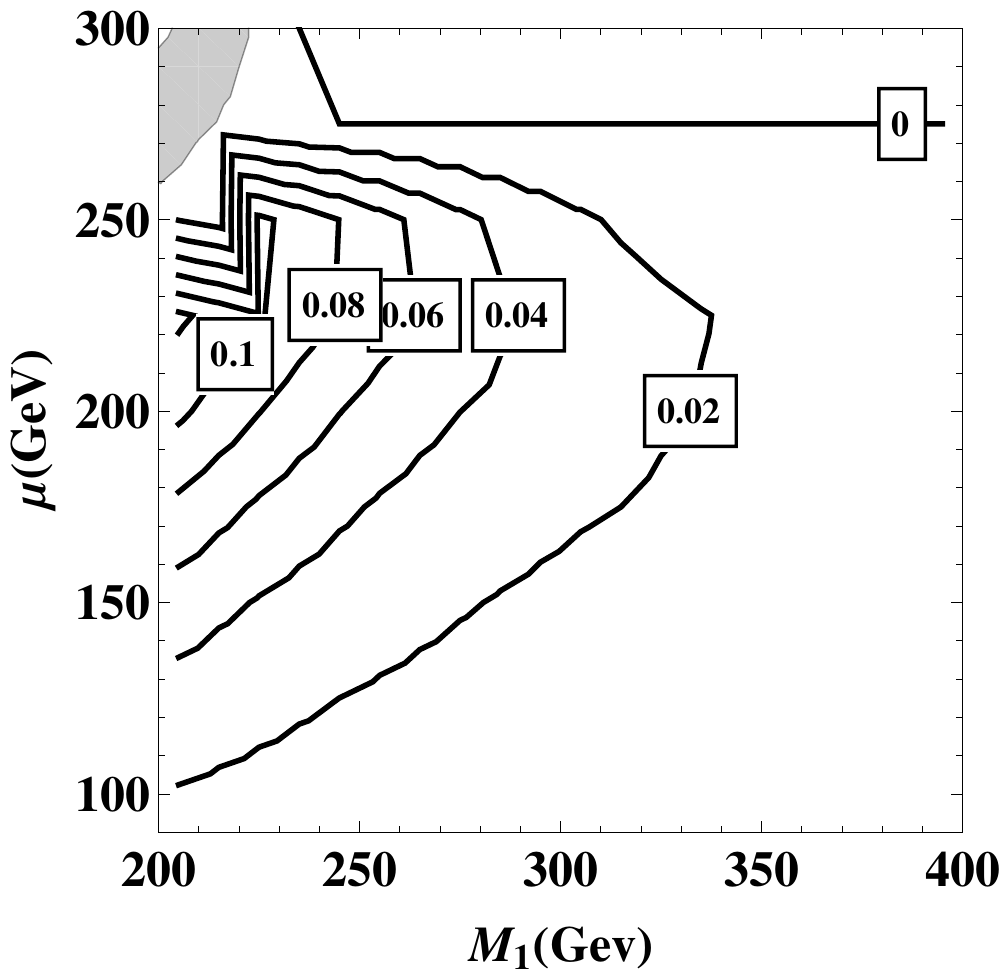}
\caption{The contours of constant branching ratio for
($ H^0 \rightarrow \tilde{\chi} ^{0}_{1} \tilde{\chi} ^{0}_{2}$)
in the $\mu- M_{1}$ for $M_A = 500$ GeV
with universal gaugino masses at the GUT scale.  The shaded region is allowed by 
WMAP data.}
\label{fig:xi1xi2_500_uni}
\end{minipage}
\end{figure}  
\begin{figure} 
\begin{center} 
\includegraphics[height=5cm,width=6cm]
{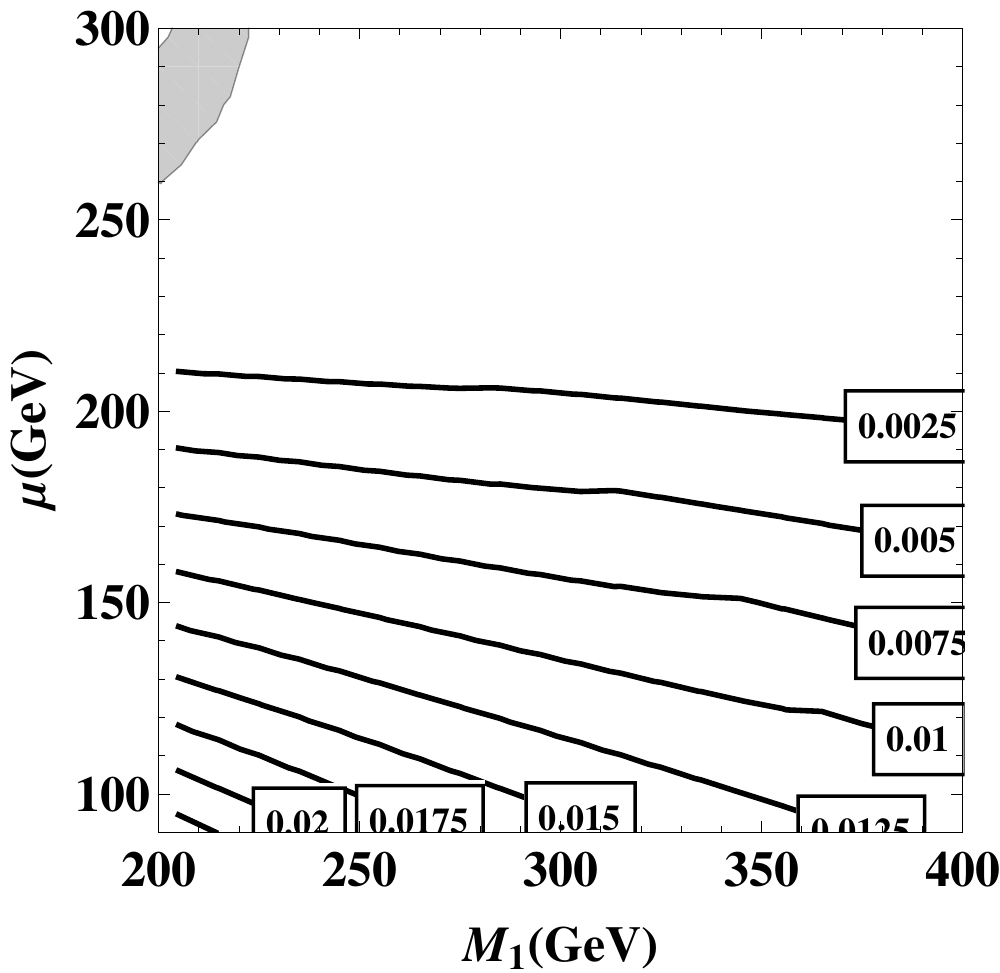} 
\caption{The contours of constant branching ratio for
($ H^0 \rightarrow \tilde{\chi} ^{0}_{2} \tilde{\chi} ^{0}_{2}$) 
in the $\mu- M_{1}$ plane for $M_A = 500$ GeV 
for universal gaugino masses at the GUT scale.  The shaded region is allowed by 
WMAP data.} 
\label{fig:xi2xi2_500_uni}
\end{center} 
\end{figure}

Having considered the case of arbitrary gaugino masses at the GUT scale 
in detail, we now turn to the case of universal boundary conditions. 
In Figs.~\ref{fig:xi1xi1_500_uni}, ~\ref{fig:xi1xi2_500_uni} and  
~\ref{fig:xi2xi2_500_uni} we have plotted the contours of constant 
branching ratio for
$ H^0 \rightarrow \tilde{\chi} ^{0}_{1} \tilde{\chi} ^{0}_{1}$ ,  
$ H^0 \rightarrow \tilde{\chi} ^{0}_{1} \tilde{\chi} ^{0}_{2},$ 
and   $ H^0 \rightarrow \tilde{\chi} ^{0}_{2} \tilde{\chi} ^{0}_{2},$  
respectively for a 500 GeV second lightest Higgs boson
with universal boundary condition on the gaugino masses at the grand unified
scale. In this case the boundary conditions  imply $M_2 \approx 2 M_1$ 
and $M_3 \approx 7 M_1$ at the electroweak scale. 
Here we can see the invisible branching ratios for
$\tilde{\chi} ^{0}_{1} \tilde{\chi} ^{0}_{1}$ can attain a value 9\% and  
$\tilde{\chi} ^{0}_{1} \tilde{\chi} ^{0}_{2}$ can attain a value 
of 10\%. The branching ratio  
$ H^0 \rightarrow \tilde{\chi} ^{0}_{2} \tilde{\chi} ^{0}_{2}$ 
can at most be 2\%. We have taken the range of $M_1$ to be 
$200$~GeV to $400$~GeV for this case. For this case
$M_2$ and $M_3$ depend on $M_1$ and to respect the experimental 
constraints we need to take $M_3$  i.e the gluino mass above 1400 GeV. 
As discussed in the previous section for values of  $\mu$ larger than
275 GeV, the  mass of the lightest neutralino exceeds the limiting 
value of $250$ GeV. Hence, a $500$~ GeV Higgs boson
cannot decay into lightest neutralino pairs. In the case of 
the second lightest neutralinos, $\mu$ = $250$~ GeV is 
the limiting value because the mass of the second lightest 
neutralino attains a value of  $250$~ GeV 
as can be seen from  Fig.~\ref{fig:M_chi2_uni}.

We note from the  behavior of the branching 
ratios that  $m_{\tilde{\chi} ^{0}_{1}}$ and 
$m_{\tilde{\chi} ^{0}_{2}}$  do not have a significant dependence
on $M_1$ but depend on  $\mu.$  This pattern is depicted in 
Figs.~\ref{fig:M_chi1_uni} and ~\ref{fig:M_chi2_uni}. 
The behavior of  
$ H^0 \rightarrow \tilde{\chi} ^{0}_{2} \tilde{\chi} ^{0}_{2}$ 
can also be understood in the same manner. We note that
$m_{\tilde{\chi} ^{0}_{2}}$ 
changes linearly with $\mu$ and in large parts of the 
parameter space it is practically equal to $\mu$. 
Hence from Eq.~\ref{Z_{2i}} 
we can see that the width  
of $ H^0 \rightarrow \tilde{\chi} ^{0}_{2} \tilde{\chi} ^{0}_{2}$ 
increases with $M_1$. We have checked that the total width 
decreases with increasing $M_1$. Hence the branching 
fraction also increases with $M_1$. The total width 
decreases with $\mu.$ Hence to get constant branching 
ratio we need increasing values of $M_1$ with decreasing values of
$\mu$.  This can be seen  from Fig.~\ref{fig:xi2xi2_500_uni}. 
The behavior of 
$ H^0 \rightarrow \tilde{\chi} ^{0}_{1} \tilde{\chi} ^{0}_{1}$ 
is also similar to that of 
$ H^0 \rightarrow \tilde{\chi} ^{0}_{2} \tilde{\chi} ^{0}_{2}.$ 
The branching fraction is dominated by  $( M_1-m_{\tilde{\chi} ^{0}_{1}})$ 
term which increases with $M_1$. We note that  for large values of
$M_1$ the behavior is almost same as  
that of $ H^0 \rightarrow \tilde{\chi} ^{0}_{2} \tilde{\chi} ^{0}_{2}.$ 
The shaded region in these Figs. represents the regions allowed by the WMAP data. 
As we can see from Figs.~\ref{fig:xi1xi1_500_uni}, ~\ref{fig:xi1xi2_500_uni} and  
~\ref{fig:xi2xi2_500_uni} the universal boundary condition scenario is very much constrained by the WMAP data. 
Only small regions of the parameter space are allowed which can give rise to some invisible branching ratio.

\begin{figure}
\begin{minipage}[b]{0.45\linewidth}
\centering
\includegraphics[width=6cm, height=5cm]{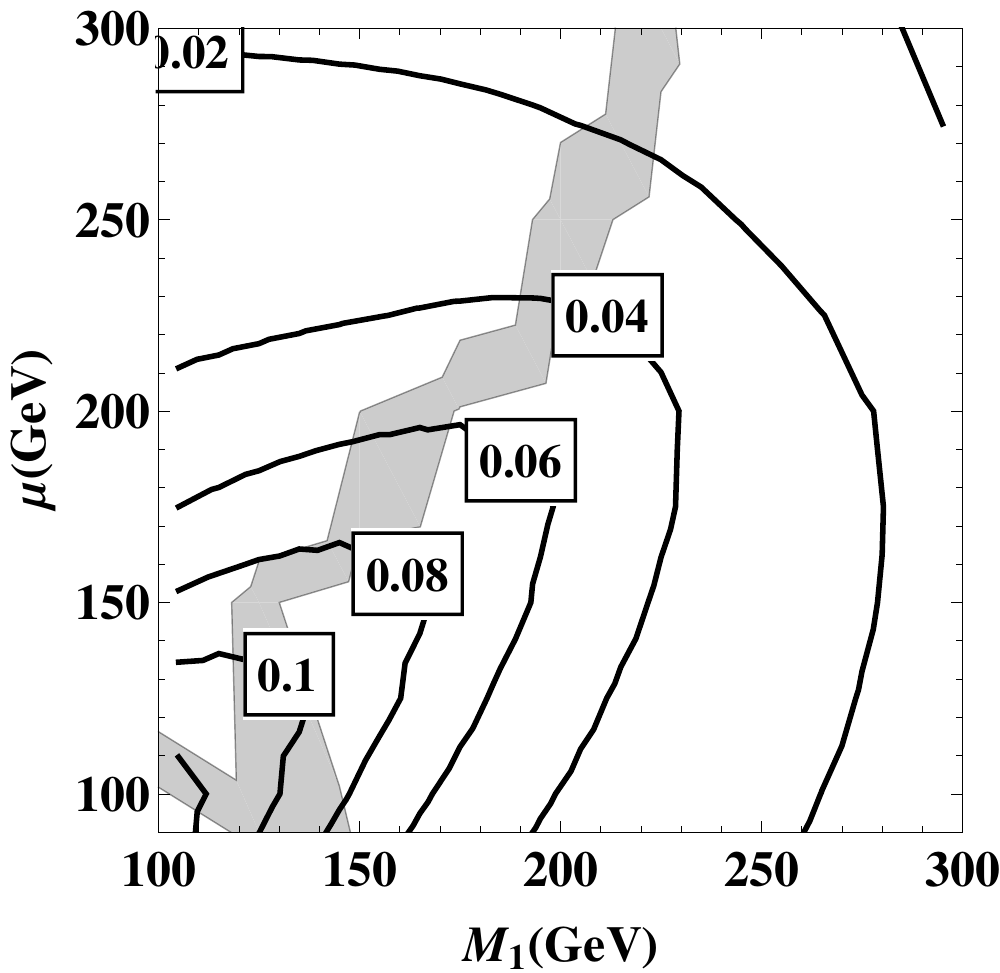}
\caption{The contours of constant branching ratio 
for  ($ H^0 \rightarrow \tilde{\chi} ^{0}_{1} \tilde{\chi} ^{0}_{1}$) 
in the $\mu- M_{1}$ plane for $M_A = 500$ GeV for
SO(10) with non-universal gaugino mass boundary condition.  The shaded region is allowed by 
WMAP data.}
\label{fig:xi1xi1_500_nonuni1}
\end{minipage}
\hspace{0.4cm}
\begin{minipage}[b]{0.45\linewidth}
\centering
\includegraphics[width=6cm, height=5cm]{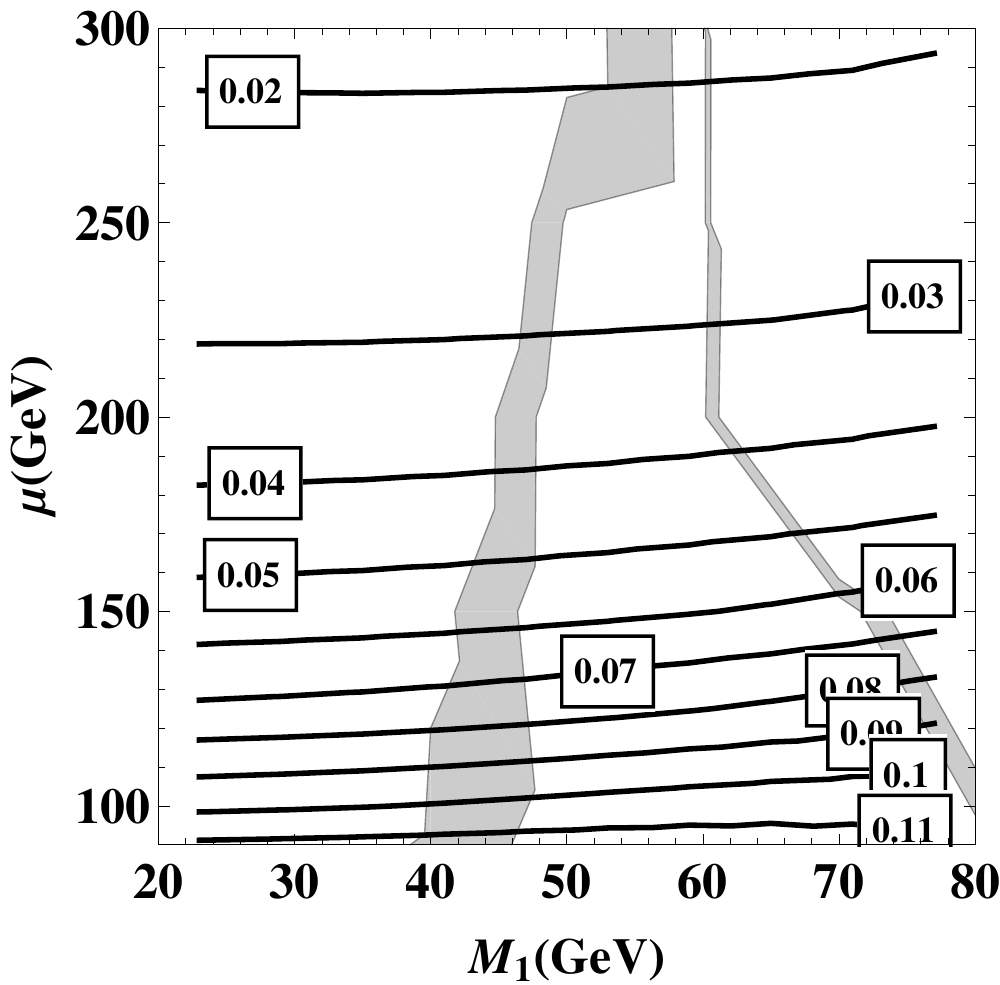}
\caption{The contours of constant branching ratio of 
($ H^0 \rightarrow \tilde{\chi} ^{0}_{1} \tilde{\chi} ^{0}_{1}$) 
in the $\mu- M_{1}$ plane for  $M_A = 500$ GeV 
for $E_6$ with non-universal gaugino mass boundary condition.  The shaded region is allowed by 
WMAP data.}
\label{fig:xi1xi1_500_nonuni2}
\end{minipage}
\end{figure}

\bigskip

In Fig.~\ref{fig:xi1xi1_500_nonuni1} we have considered an example 
of non-universal boundary condition on the gaugino masses. 
For this we have chosen a $\bf 210$ dimensional representation of 
the gauge group SO(10) in which $M_1:M_2:M_3$ is in the ratio $1:6:-14.3$. 
We have plotted the contours of constant branching ratio for
$ H^0 \rightarrow \tilde{\chi} ^{0}_{1} \tilde{\chi} ^{0}_{1}$  
in the $\mu-M_1$ plane. We can see that the branching ratio can go 
upto 10\% for a small region of the parameter space. 
The shaded region in this Fig. represents the region allowed by the WMAP data.

In Fig.~\ref{fig:xi1xi1_500_nonuni2} we have considered another 
example of non-universal boundary condition on gaugino masses
for a $\bf 2430$ dimensional representation of the gauge group $E_6$. 
The gaugino masses are in the ratio $M_1:M_2:M_3$ = $1:50.2:70.9$. 
This particular representation is somewhat special. Because out of all 
the non-universal boundary conditions this is the only representation 
in which r(defined as the ratio $M_1/M_2$) satisfies the the condition 
$r \leq 0.04,$ resulting in a light neutralino.
This actually allows the 126 GeV Higgs to decay into lightest neutralinos.
In this case we can see the branching ratio for
$ H^0 \rightarrow \tilde{\chi} ^{0}_{1} \tilde{\chi} ^{0}_{1}$ 
can go upto 11\% for a tiny region of the $\mu-M_1$ parameter space. 
The shaded region in this Fig. represents the region allowed by the WMAP data.

  For the case of Fig.~\ref{fig:xi1xi1_500_nonuni1} we see that for a 
particular value of $M_1$ the branching fraction decreases with $\mu$. 
Because the quantity $M_1-m_{\tilde{\chi} ^{0}_{1}}$ decreases with 
$\mu$ at a particular value of $M_1$, $m_{\tilde{\chi} ^{0}_{1}}$ increases 
with $\mu$ as can be seen from Fig.~\ref{fig:M_chi1_nonuni1}. 
Hence $Z_{13}$ and $Z_{14}$ decrease with $\mu$. 
The term $Z_{12}$ also decreases with $\mu$, because the denominator 
$6 M_1-m_{\tilde{\chi} ^{0}_{1}}$  decreases with $\mu,$ 
but the rate of decrease is lower than the rate of increase 
of  $M_1-m_{\tilde{\chi} ^{0}_{1}}$ in the numerator. 
The total width decreases with $\mu$, but here also the rate of 
decrease is less, hence the branching fraction decreases 
with $\mu$ for a particular value of $M_1$. 

 Fig.~\ref{fig:xi1xi1_500_nonuni2} corresponds to the 
situation  where the mass of the lightest neutralino is almost equal 
to $M_1$ for all $\mu,$ as can be seen from Fig.~\ref{fig:M_chi1_nonuni2}. 
Hence $(M_1-m_{\tilde{\chi} ^{0}_{1}})$ is a very small quantity which is 
more or less constant as a function of  $\mu$. 
Hence $Z_{12},Z_{13}$ and $Z_{14}$ all decrease with $M_1$ for a 
particular $\mu$, having $m_{\tilde{\chi} ^{0}_{1}},$ which is equal to $M_1,$ 
in the denominator. But the total width also decreases with $M_1$. 
Hence the branching fraction remains almost constant as a function of $M_1$.

\bigskip

It is known that
the 126 GeV Higgs boson of the MSSM can decay invisibly  
only for arbitrary gaugino masses at the GUT scale.
Here we find that for the situation for the
second lightest Higgs boson MSSM is far more interesting, 
because it can decay  invisibly, not only for arbitrary gaugino masses 
but also for constrained  boundary conditions of 
universal and non-universal gaugino masses at the GUT scale. 

  From the plots we notice that there are kinks in all the contour plots 
for branching ratios  studied in this paper. Although it is  
difficult to obtain analytical expressions for the constant branching ratio contours, 
one thing that can be easily understood
is that depending on the mass of the neutralinos different 
decay channels will open up for different range of values of 
$\mu$ and $M_{1}$. Whenever the total decay width in the denominator changes because 
of new channels opening up, the constant branching ratio contour plots show  kinks.

   The range of values of $M_{1}$ and $\mu$ considered in this
paper is adequate for our study, because beyond this range 
of $M_{1}$ and $\mu$ the branching ratios fall and are therefore
not relevant.

\bigskip

\begin{figure}
\begin{minipage}[b]{0.45\linewidth}
\centering
\includegraphics[width=6cm, height=5cm]{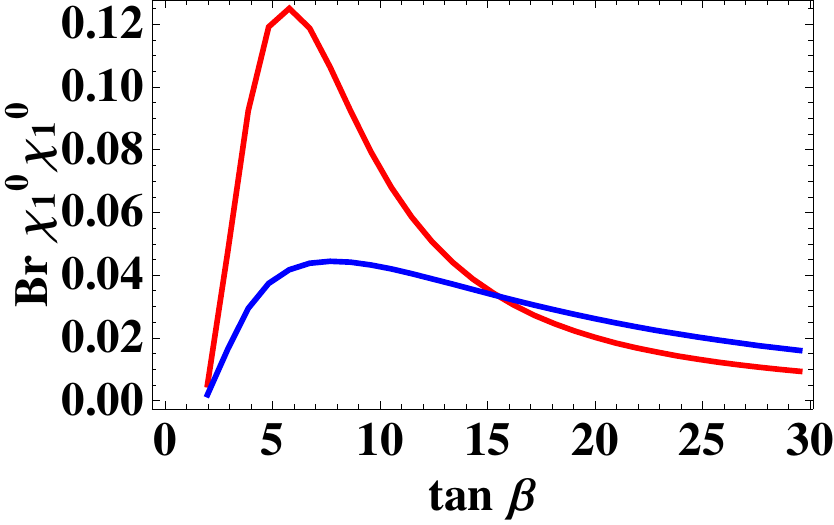}
\caption{Branching ratio of 
($ H^0 \rightarrow \tilde{\chi} ^{0}_{1} \tilde{\chi} ^{0}_{1}$) 
with $M_H$ = 300~GeV(red),500~GeV(blue) as a function of $\tan \beta$.}
\label{fig:xi1xi1_tanbeta}
\end{minipage}
\hspace{0.4cm}
\begin{minipage}[b]{0.45\linewidth}
\centering
\includegraphics[width=6cm, height=5cm]{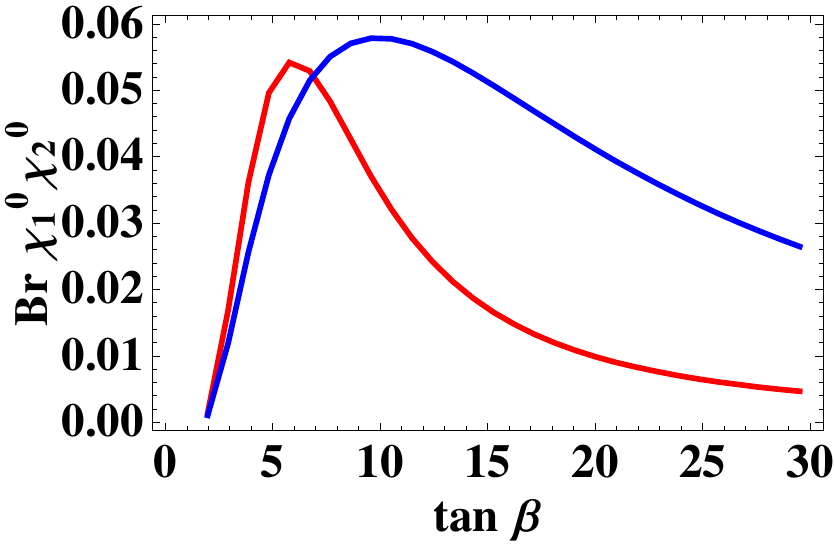}
\caption{Branching ratio of 
($ H^0 \rightarrow \tilde{\chi} ^{0}_{1} \tilde{\chi} ^{0}_{2}$) 
with $M_H$ = 300~GeV(red),500~GeV(blue) as a function of $\tan \beta$.}
\label{fig:xi1xi2_tanbeta}
\end{minipage}
\end{figure}  

  \begin{figure} 
\begin{center} 
\includegraphics[height=5cm,width=6cm]{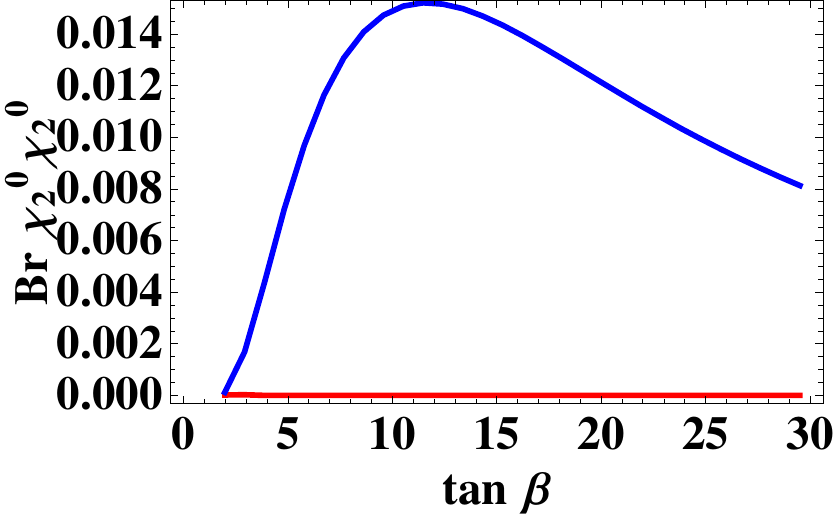} 
\caption{Branching ratio of 
($ H^0 \rightarrow \tilde{\chi} ^{0}_{2} \tilde{\chi} ^{0}_{2}$) 
with $M_H$ = 300~GeV(red),500~GeV(blue) as a function of $\tan \beta$.} 
\label{fig:xi2xi2_tanbeta}
\end{center} 
\end{figure}

\bigskip

 We have also considered the dependence of the branching ratios
for the decays
$ H^0 \rightarrow \tilde{\chi} ^{0}_{1} \tilde{\chi} ^{0}_{1}$ ,$ H^0 \rightarrow \tilde{\chi} ^{0}_{1} \tilde{\chi} ^{0}_{2}$ and $ H^0 \rightarrow \tilde{\chi} ^{0}_{2} \tilde{\chi} ^{0}_{2}$
on $\tan \beta.$  Figs.~\ref{fig:xi1xi1_tanbeta},~\ref{fig:xi1xi2_tanbeta}
and ~\ref{fig:xi2xi2_tanbeta} show the 
dependence of these two branching ratios 
on $\tan \beta$ for $M_{A} \approx M_H$= 300 and 500 GeV
for the choice of parameters shown in Table~\ref{tab:tanbetaparam}.
These values of parameters are consistent with all
current experimental constraints.
\begin{table}[htb]
\renewcommand{\arraystretch}{1.0}
\begin{center}
\vspace{0.5cm}
\begin{tabular}{|c|c|c|c|}
\hline
$M_1$ = 120 GeV &$M_S$ = 1.5 TeV &$M_A$ = 300,500 GeV 
&$M_2$ = 200 GeV \\
\hline
$M_3$ = 1402 GeV    &$A_t$= 3600 GeV &$A_b$= 3600 GeV &$\mu$= 200 GeV \\
\hline 
\end{tabular}
\end{center}
\vspace{-0.5cm}
\caption{Input parameters for the plots of branching ratios 
versus $\tan \beta.$}
\renewcommand{\arraystretch}{1.0}
\label{tab:tanbetaparam}
\end{table}
\noindent
  
 The behavior of the $\tan \beta$ dependence of the invisible branching ratios
can be understood in the following manner. 
In MSSM, the  $H^0 \rightarrow b \bar b$ coupling 
is $ \frac{ \cos \alpha}{\cos \beta }$ times its SM value,
and $H^0 \rightarrow t \bar t$ coupling is $\frac{ \sin \alpha}{\sin \beta}$ 
times the SM value.

 At low values of $\tan \beta$ the 
$H^0$ coupling to the up-type quarks is large, 
hence in this region its decay to top-antitop quarks is dominant, 
if kinematically allowed. If the  second lightest Higgs has a mass of
500 GeV, the branching ratio  for this decay 
is 57\%. In this region, the decay channel $b \bar b$ or 
the invisible decay channels are not significant.
If the Higgs is not so heavy~(around $300$~GeV), 
then top quark channel is kinematically not available, and the main decay channel then 
will be to two lightest Higgs and some of the invisible channels.

   In case of moderate $\tan \beta$, say $\tan \beta = 7$, 
invisible branching ratio for 
$H^0 \rightarrow \tilde{\chi} ^{0}_{1} \tilde{\chi} ^{0}_{1}$ 
peaks for both $300$~GeV as well as $500$~Gev Higgs boson.

The branching ratios for $H^0 \rightarrow \tilde{\chi} ^{0}_{1} \tilde{\chi} ^{0}_{2}$ also peaks at moderate values of $\tan \beta$ for $300$ and $500$ GeV second lightest Higgs. On the other hand $\tilde{\chi} ^{0}_{2}$ being very heavy, the $300$ GeV second lightest Higgs cannot decay into two $\tilde{\chi} ^{0}_{2}$s for any $\tan \beta$ for the parameter space considered. For the $500$ GeV second lightest Higgs the branching ratio for $H^0 \rightarrow \tilde{\chi} ^{0}_{2} \tilde{\chi} ^{0}_{2}$ peaks at moderate values of $\tan \beta$.

   For large values of $\tan \beta \approx 30$, 
Higgs coupling to the down-type quarks is dominant,
and $b \bar b$ becomes the dominant decay mode. 
The invisible decay channels are once again  insignificant. 
   
   In the case of lightest Higgs decay, in the low $\tan \beta$ region, 
the invisible decay would be significant as the top decay channel 
is kinematically closed.
In the case of the second lightest Higgs boson, low values of
$\tan \beta$ do not give significant invisible decay width, 
but for moderate values of $\tan \beta \approx 10$ the invisible
branching ratio can be significant.

\bigskip

  \begin{figure} 
\begin{center} 
\includegraphics[height=6cm,width=8cm]{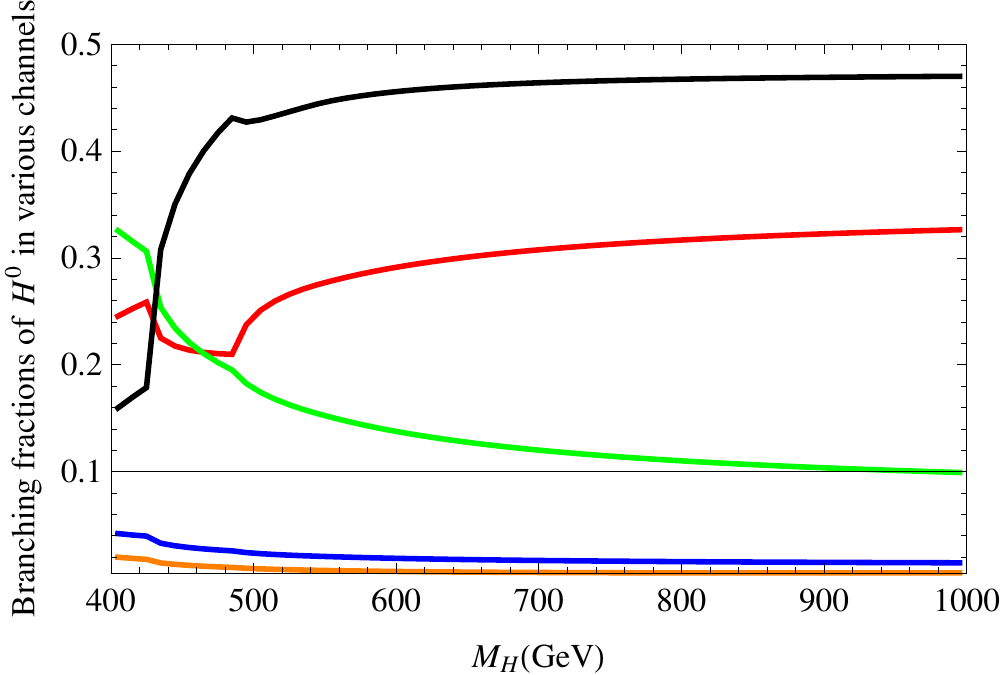} 
\caption{The branching fraction of the heavier Higgs to 
$b \bar b$(green), neutralinos(red), charginos(black), $\tau \tau$ (blue), 
lightest Higgs(orange) as a function of $M_H$.} 
\label{fig:figure}
\end{center} 
\end{figure}

In the Fig.~\ref{fig:figure} we have considered the 
branching fraction of heavier Higgs boson
to all the dominant decay channels. The dominant visible channels 
are $b \bar b, \tau \bar \tau, hh$~\cite{CMS-TDR-2006,Carena:2014ac}. 
With increase in the values of $M_H,$ 
the branching fraction for the $b \bar b$ channel decreases, 
$\tau \bar \tau, hh$ channels are not significant,
and chargino and neutralino channels become important.
Neutralino branching fraction can be as large as
30\%, and chargino branching fraction can be about 50 \%. 
For this study the input parameters are $M_1=150$, $M_2=200$, $M_3=1400$, 
$\mu =200$ and $\tan \beta = 10$. From this study 
we can see that if the second lightest Higgs is heavy enough it can 
have major decay channels into the electroweak-inos. It can be inferred that in 
the future LHC run during the direct 
search for the second lightest Higgs boson, its decay into 
supersymmetric particles may  play an  important role 
in the decoupling regime.



\subsection{``Quasi-invisible" decays}

We would like to draw attention of the reader to the fact that all the decay channels of the second lightest Higgs boson that we have considered and studied in detail in this paper are not truly invisible. $H^0 \rightarrow \tilde{\chi} ^{0}_{1} \tilde{\chi} ^{0}_{1}$ channel is truly invisible as the lightest neutralino $\tilde{\chi} ^{0}_{1}$ is the lightest supersymmetric particle. It is the only stable supersymmetric particle assuming conservation of R-parity. All other heavy neutralinos will decay into the LSP and SM particles. Hence  $H^0 \rightarrow \tilde{\chi} ^{0}_{1} \tilde{\chi} ^{0}_{2}$ and  $H^0 \rightarrow \tilde{\chi} ^{0}_{2} \tilde{\chi} ^{0}_{2}$ cannot be considered invisible. In  ref~\cite{Han:2013ac} the authors have extensively studied the electrowikino decays for different regions of the parameter space of the electroweak sector without assuming any SUSY-breaking mediation scenario. There is a region of the parameter space defined as scenario C  in the paper where $|\mu| < M_1,M_2$. In this region of the parameter space $\tilde{\chi} ^{0}_{1}$ and $\tilde{\chi} ^{0}_{2}$ are both Higgsino like and their masses are almost equal to the Higgsino mass $\mu$ and hence they are almost mass-degenerate. It has been pointed out in the same reference that the LSP multiplet production will be difficult to observe at the LHC because of the mass degeneracy and the soft decay products especially when the production is suppressed. Hence one has to be extremely careful while looking at the decay of the electrowikinos at the future LHC. Where the  $\tilde{\chi} ^{0}_{1}$ and $\tilde{\chi} ^{0}_{2}$ are almost mass degenerate, we have termed this kind of decays $H^0 \rightarrow \tilde{\chi} ^{0}_{1} \tilde{\chi} ^{0}_{2}$ and  $H^0 \rightarrow \tilde{\chi} ^{0}_{2} \tilde{\chi} ^{0}_{2}$ as 'quasi-invisible' decay. It is worth mentioning that, one must expect that the International Linear Collider(ILC) will be able to identify the soft decay products, leptons and jets of mass $10$ GeV or less, which are produced in the mass-degenerate case, because ILC has a cleaner environment for event reconstruction~\cite{Han:2013ac}.

Another intriguing possibility that could render some of these channels ``visible" is to adapt a strategy that has been used in the past for dark matter searches. In other words a promising search for invisible decays is in the monojet channel as in the case for dark matter search at the LHC.
When the second lightest Higgs boson decays into invisible particles, monojet searches
involving initial state radiation from the gluons that fuse to produce the Higgs boson could be used
in these channels so the the Higgs decaying into light neutralinos does not necessarily escape the detection at the colliders. This same procedure is used to look for dark matter at the LHC. In case of invisible decays one has to look for monojet signals $H^0 \rightarrow \tilde{\chi} ^{0}_{1} \tilde{\chi} ^{0}_{1} + jet$ as $\tilde{\chi} ^{0}_{1}$ is the lightest supersymmetric particle .
In ref~\cite{Djouadi:2013ab,bai:2012,Englert:2012} it has been shown that the monojet
signature carries a good potential to constrain the invisible
decay width of a $\approx$ 125GeV Higgs boson in a model independent
fashion using the monojet search results by ATLAS and CMS~\cite{CMS:2012,ATLAS:2012}. Now in case of MSSM with two CP-even Higgs, we can employ a different strategy while looking for the second lightest Higgs boson. One can calculate the production cross section $\times$ invisible branching in the MSSM for specific regions of the parameter space. There are several Standard Model processes which can act as background for the monojet signals. $pp \rightarrow Z(\rightarrow \nu \bar \nu)+jet$ is the main irreducible background with the same topology as our signal. There is QCD background as well. The backgrounds can be estimated~\cite{Han:2014}. The background can be reduced significantly using several $p_T$ cuts on the jet and missing transverse energy cut. Then one can find the dependence of signal significance $\cal{S}$=$ N_S/ \sqrt{N_B}$($N_S$ is the number of signal events and $N_B$ is the number of background events) on different parameters for 14 TeV HL-LHC with the desired integrated luminosity $\cal{L}$. 
One can use the  LHC monojet search results at 14 TeV, ie. the limits on the monojet events to probe the regions in the MSSM parameter space spanned by $M_A, \tan \beta, \mu$~\cite{Biplob:2014}. Thus one can exclude regions of the MSSM parameter space at a desired signal significance(90 \% or 95 \% C.L).

\section{Discussion and conclusion}

In this work we have considered the possibility of invisible decays
of the second lightest Higgs boson$(H^0)$ in the MSSM in the 
decoupling regime.  In the past, various studies have 
shown that certain regions of the parameter space of MSSM, allow a Higgs
boson in the mass range 123-129 GeV both in the decoupling and non-decoupling regime, satisfying the LHC constraints.
For most of the parameter space, the lightest Higgs decay to 
the lightest neutralinos is kinematically allowed,
leading to invisible decay modes. 
The main objective of those works was to prove that it would therefore 
be very important to study the couplings of the newly
discovered particle at high precision. Global fits have been 
performed on the couplings of
the newly discovered particle, 
in order to place upper bounds on the invisible decay width. 
Taking into account
these bounds, the parameter space of these new physics scenarios can be further constrained, since the regions
giving a large invisible Higgs decay branching ratio will be in 
conflict with the experiments.

In the present work, which is a sequel to previous
investigations ~\cite{Ananthanarayan:2013fga}, we are looking at the 
problem from a different point of view. We are asking a related 
question.  We have considered the intriguing possibility that 
the heavier CP even Higgs$(H^0)$ of the MSSM would decay into invisible LSP's.
Knowing the invisible branching ratios of the second lightest Higgs boson one can look for it
in the future  HL-LHC or ILC through the monojet signals. From the limits given by the colliders on the
monojet signal one can then probe the MSSM parameter space at a desired confidence level. Hence it will also
be possible to detect or rule out the second lightest Higgs boson in the MSSM in a certain mass range from the monojet searches for its invisible decay. In other words, in our previous 
paper~\cite{Ananthanarayan:2013fga}, the study has been done 
to constrain the MSSM parameter space looking at the possible 
invisible decay width of the newly discovered 126 GeV Higgs 
boson. But in this work we are following a different route 
and trying to make a theoretical prediction on the search 
for the second lightest Higgs boson of the MSSM.
Our point is to understand the invisible decay width of $H^{0}$ consistent
with the current experimental constraints.
 In the scenarios that we considered earlier, which
partly motivated the present study, we had considered that the 125 GeV Higgs boson
had some partial decay channels into light neutralinos in the MSSM, or that
it would decay into neutralinos or light CP odd Higgs particles which could
be present in the spectrum of the NMSSM.  
In ~\cite{ Pandita:2014nsa} the intriguing possibility
was considered that the 125 GeV Higgs was not that lightest CP even Higgs
but the heavier one, while the lightest evaded detection altogether.  Here, we
have more conservative assumptions, viz., that the 125 GeV resonance is
indeed the lightest CP even Higgs, but consider the possibility that the
heavier  one has invisible decays.  This scenario can arise naturally in the
decoupling region, i.e., the second lightest Higgs boson is 
quite heavy and could be produced
at future experiments at the LHC which will eventually decay into lighter particles.
These final states include neutralinos, charginos, $ bb,ZZ, \tau \tau, hh$.  In the present work
we have tried to carry out an exhaustive search of the MSSM parameter space
where the second lightest Higgs boson can decay into the lightest neutralinos and therefore remains invisible.
We have considered the decay of the second lightest Higgs into the second lightest neutralinos also, because there are certain regions of the parameter space where the lightest and the second lightest neutralino are almost mass-degenerate and hence the second lightest neutralino may remain invisible(`quasi-invisible') at the LHC.
  We have used semi-analytical
formulas as a guide to our study.
In the decoupling regime, which is a realistic scenario,
the values of  $M_A$ and $M_H$ are nearly equal.

For our parameter choices guided by recent constraints by LEP and LHC,
we have taken the gaugino masses at the electroweak scale to be 
within present constraints. We have chosen $A_t$ judiciously to
get the lightest Higgs mass in the range in which it has been detected. We have taken soft SUSY breaking scale $M_S$ to large, around 1.5 TeV consistently.
In this paper, we have presented a detailed analysis first of the Higgs sector
as well as that of the neutralino sector, in order to isolate 
the conditons under which it becomes kinematically feasible for $(H^0)$ to
decay invisibly. We have considered three different scenarios and have scanned the parameter space in terms of invisible decay of the second lightest Higgs boson in these three scenarios, namely arbitrary, universal and non-universal boundary condition on the soft gaugino masses at the GUT scale.
 Our analysis reveals that it
is typically not possible to have such regions in the MSSM with
GUT scale  universal boundary conditions on the soft gaugino masses.
This can be seen from Figs.~\ref{fig:xi1xi1_500_uni}, 
~\ref{fig:xi1xi2_500_uni} and ~\ref{fig:xi2xi2_500_uni}.
The situation remains similar even with non-universal boundary conditions as
can be seen from Figs.~\ref{fig:xi1xi1_500_nonuni1} 
and ~\ref{fig:xi1xi1_500_nonuni2}.  The main reason for this is that there is
not sufficient freedom in the choice of the gaugino masses with these
boundary conditions.  On the other hand,
relaxing all constraints, as is the case with general (arbitrary) 
boundary conditions,
allows the invisible branching ratios to be considerable.  
This can be seen from
Figs.~\ref{fig:xi1xi1_300},~\ref{fig:xi1xi1_500},~\ref{fig:xi1xi2_300},
~\ref{fig:xi1xi2_500},~\ref{fig:xi2xi2_300} and ~\ref{fig:xi2xi2_500}.  

  From our study we conclude that there is a significant portion of 
the parameter space where the invisible decays
can be quite significant.  For instance, with universal 
or non-universal boundary conditions,
the invisible branching ratio is not enhanced, but if we relax this constraint
we can have more significant branching ratio in invisible decays.
Hence the monojet searches for the invisible decay of the second lightest Higgs boson at the future colliders 
can be used as a probe to look for the second lightest Higgs boson as well as to put constraints on the regions of
the MSSM parameter space. This would be most useful in the case of arbitrary boundary condition on the gaugino masses at the GUT scale. One has to be more careful and find different strategies in the case of universal and non-universal boundary condition.

\section{acknowledgements} P.N.P would like to thank 
Inter-University Centre for Astronomy and 
Astrophysics, Pune, where part of this work was done, for hospitality. The 
work of P.N.P is supported by the Raja Ramanna Fellowship of the Department of
Atomic Energy, and
partly by the J. C. Bose National Fellowship 
of the Department of 
Science and Technology, India.
J.L would like to thank Dr. Biplob Bhattacherjee and Dr. Monalisa Patra for useful and enlightening
discussions. B.A would like to thank Professor Daniel Wyler and the University of Zurich, where part of this work was done,  for hospitality .

 

\end{document}